\newcommand{\be}{\begin{equation}}
\newcommand{\ee}{\end{equation}}
\newcommand{\brr}{\begin{eqnarray}}
\newcommand{\err}{\end{eqnarray}}
\newcommand{\nn}{\nonumber}
\newcommand{\bd}{\begin{displaymath}}
\newcommand{\ed}{\end{displaymath}}

\newcommand{\bib}{\bibitem}
\newcommand{\bfig}{\begin{figure}}
\newcommand{\efig}{\end{figure}}
\def\alf{\alpha}
\def\bet{\beta}

\def\om{\omega}
\def\eps{\epsilon}
\def\rpar{\right)}
\def\lpar{\left(}
\def\rbk{\right]}
\def\lbk{\left[}
\def\rbr{\right\}}
\def\lbr{\left\{}
\def\lb{\label}
\def\im{{\rm i}}
\def\ro{\mbox{\boldmath $\rho$}}
\def\opb{\mbox{\boldmath $\beta$}}
\def\opf{\mbox{\boldmath $\varphi$}}
\def\sig{\mbox{\boldmath $\sigma$}}
\def\ene{\mbox{\tiny {\rm N}}}
\def\dab{\mbox{\tiny {\rm W}}}
\def\ipa{\mbox{\tiny {\rm A}}}
\def\ief{\mbox{\tiny {\rm eff}}}
\def\opu{\mbox{\boldmath ${\rm U}$}}
\def\ima{\mbox{${\rm Im}$}}

\def\rg{\rangle}
\def\llg{\langle}
\def\hp{\hspace{\parindent}}
\def\Tr{{\rm Tr}}
\def\half{\frac{1}{2}}
\documentclass[pra,aps]{revtex4}
\usepackage{ifvtex}
\usepackage{ifpdf}
\usepackage[dvips]{graphicx}
\usepackage{upmath}
\usepackage{amssymb}
\usepackage{amsmath}
\usepackage{eucal}
\usepackage[all]{xy}
\begin{document}
%
\title{Quantitative aspects of entanglement in the driven Jaynes-Cummings model\footnote{\bf Journal of Modern Optics 53 (18), 2733
(2006)}}
\author{Marcelo A. Marchiolli}
\affiliation{Instituto de F\'{\i}sica Te\'{o}rica, Universidade Estadual Paulista, \\
             Rua Pamplona 145, 01405-900 S\~{a}o Paulo, SP, Brazil \\
             Email: marcelo$\_$march@bol.com.br}
\date{\today}
%
\begin{abstract}
\vspace*{0.1mm}
\begin{center}
\rule[0.1in]{142mm}{0.4mm}
\end{center}
Adopting the framework of the Jaynes-Cummings model with an external quantum field, we obtain exact analytical expressions of the
normally ordered moments for any kind of cavity and driving fields. Such analytical results are expressed in the integral form, with
their integrands having a commom term that describes the product of the Glauber-Sudarshan quasiprobability distribution functions for
each field, and a kernel responsible for the entanglement. Considering a specific initial state of the tripartite system, the normally
ordered moments are then applied to investigate not only the squeezing effect and the nonlocal correlation measure based on the total
variance of a pair of Einstein-Podolsky-Rosen type operators for continuous variable systems, but also the Shchukin-Vogel criterion.
This kind of numerical investigation constitutes the first quantitative characterization of the entanglement properties for the driven
Jaynes-Cummings model.\bigskip \\
keywords: Driven Jaynes-Cummings Model, Entanglement, Inseparability Criteria \\
PACS: 03.65.-w; 03.65.Ud; 03.67.Mn \\
\vspace*{0.1mm}
\begin{center}
\rule[0.1in]{142mm}{0.4mm}
\end{center}
\end{abstract}
\maketitle
\section{Introduction}
%
\hp Entanglement and nonlocal correlations are abstract concepts that naturally appear in quantum mechanics when the superposition 
principle is applied to composite systems. Nowadays, these concepts play an essential role not only in the quantum computation scenario
\cite{r1} and quantum information theory \cite{r2}, but also in the context of relativity theory \cite{r3}. For instance, one of
the main tasks of quantum information theory (QIT) is to develop a quantitative characterization of the entanglement properties and
quantum correlation effects for multipartite physical systems described by continuous and/or discrete variables \cite{r4,r5}. In this
sense, recent theoretical and experimental developments in QIT, which are based on the continuous variable regime and with emphasis on
quantum optical implementations involving quadrature amplitudes of the electromagnetic field, have appeared in the literature 
\cite{r6,r7,r8,r9,r10,r11,r12,r13,r14,r15,r16,r17,r18,r19,r20,r21,r22}. 

From the theoretical point of view, Duan {\it et al.} \cite{r6} proposed a simple inseparability criterion based on the calculation of
the total variance of a pair of Einstein-Podolsky-Rosen (EPR) type operators for continuous variable states. Basically, this criterion
states that `for any separable continuous variable states, the total variance is bounded from below by a certain value resulting from
the uncertainty relation, whereas for entangled states this bound can be exceeded.' Consequently, the violation of this bound can be
interpreted as `a sufficient condition for inseparability of the states.' Similarly, Simon \cite{r7} adopted the Peres-Horodecki 
criterion of positivity under partial transpose in the context of separability of bipartite continuous variable states, and showed that
this mathematical operation admits a geometric interpretation as mirror reflection in phase space. Thus, exploiting the consequences of
such geometric interpretation, the author derived `uncertainty principles stronger than the traditional ones, to be obeyed by all
separable states.' Following the same mathematical approach adopted in \cite{r6}, Giedke {\it et al.} \cite{r12} showed that two-mode
squeezed states maximize EPR-like correlations for a fixed amount of entanglement; besides, this result was used to determine the
entanglement of formation for all symmetric gaussian states describing bipartite systems. More recently, Shchukin and Vogel \cite{r16}
established an important set of inseparability criteria for bipartite quantum states which is formulated in terms of observable moments
associated with a variety of quantum states. To this end, the authors derived a hierarchy of necessary and sufficient conditions for
the negativity of the partial transposition of bipartite quantum states (or sufficient conditions for entanglement) that generalizes
some previous purposes from literature.

Now, from the experimental point of view, Josse {\it et al.} \cite{r14} produced in laboratory both quadrature and polarization
entanglement via the interaction between a linearly polarized coherent field and a cloud of cold cesium atoms placed in a high finesse
optical cavity. To demonstrate continuous variable entanglement in this beautiful experiment, the authors used the theoretical tools
established in \cite{r6,r12} -- namely, the inseparability criterion and the entanglement of formation -- with great success in the
experimental data analysis. Furthermore, this experiment opened new windows of investigation in similar physical systems which describe
the matter-field interactions. A feasible physical system to generate continuous variable entanglement is given by the Jaynes-Cummings
model (JCM), such theoretical tool being typically realized in cavity-QED experiments involving Rydberg atoms crossing superconducting
cavities in different frequency regimes and configurations, with relaxation rates small and well understood \cite{r23}. Beyond these
fundamental features, it is worth mentioning that multipartite gaussian states have potential applications for quantum teleportation
\cite{r24,r25} and quantum criptography \cite{r26}.

Many authors have investigated the two-mode and driven JCM in different theoretical contexts and predicted new interesting results 
(for example, see \cite{r27,r28,r29,r30,r31,r32,r33,r34,r35,r36,r37,r38,r39,r40}). However, a quantitative characterization of the
entanglement properties and a detailed analysis of the nonlocal correlation effects among the constituents (fields and atoms) of
these similar physical systems have not frequently appeared in the literature, and represent two important additional tools in the
comprehension of the atom-photon interactions which deserve to be carefully studied. Thus, the main aim of this work is to present the
first quantitative characterization of the entanglement properties for the driven JCM, which is based upon the inseparability criteria
for continuous variable systems developed in \cite{r6,r16}. To this end, we assume the atom is initially prepared in the excited state,
and the cavity and external fields are in the diagonal representation of coherent states. Next, we use the mathematical procedure
developed in \cite{r40} to obtain exact analytical expressions of the normally ordered moments for any kind of cavity and driving
fields. These analytical results are then expressed at the integral form with their integrands presenting a common term that describes
the product of the Glauber-Sudarshan quasiprobability distribution functions for each field, and a kernel responsible for the nonlocal
correlations among the constituents of the tripartite system. To illustrate our results we fix both the cavity and driving fields in a
coherent state. In particular, we attain new results within which some of them deserve to be mentioned: (i) we establish a link between
squeezing effect and entanglement for different values of detuning frequency in the driven JCM; (ii) we show that EPR uncertainty can
not be considered the first quantitative characterization of the entanglement properties of the system under investigation; and finally,
(iii) we present a numerical evidence that supports the hierarchy of necessary and sufficient conditions for inseparability derived by
Shchukin and Vogel \cite{r16}.

This work is organized as follows. Section II describes the mathematical procedure used in the derivation of exact analytical
expressions of the normally ordered moments for the driven JCM. In Section III we apply our results in the analysis of the squeezing
effect for each field fixing, for convenience, both the cavity and driving fields in a coherent state. Moreover, we also analyse (by
means of a numerical investigation) some recent proposals of inseparability criteria for physical systems described by continuous
variables (in particular, we consider the EPR uncertainty and the Shchukin-Vogel criterion). Section IV contains our conclusions.
Finally, Appendix A shows the main steps to calculate the exact analytical expressions for the generalized moments.

\section{Derivation of the normally ordered moments}

\hp In many textbooks on quantum optics \cite{r41}, the normally ordered moments are generally defined in terms of an auxiliary function
(also denominated as the normal characteristic function) which describes the normal ordering of creation and annihilation operators of
the electromagnetic field -- i.e., 
\bd
\Lambda_{\ene}^{({\rm c})}(\xi,\xi^{\ast};t) \equiv \Tr \lbk \ro(t) \, {\rm e}^{\xi {\bf c}^{\dagger}} {\rm e}^{- \xi^{\ast} {\bf c}}
\rbk \; , 
\ed
$\ro(t)$ being the density operator of the system under investigation. The connection between normally ordered moments and normal
characteristic function is promptly established through the relation
\be
\lb{mom1}
\llg {\bf c}^{\dagger r}(t) {\bf c}^{s}(t) \rg \equiv \Tr \lbk \ro(t) \, {\bf c}^{\dagger r} {\bf c}^{s} \rbk = (-1)^{s} \left.
\frac{\upartial^{r+s}}{\upartial \xi^{r} \upartial \xi^{\ast s}} \Lambda_{\ene}^{({\rm c})}(\xi,\xi^{\ast};t) 
\right|_{\xi,\xi^{\ast} = 0}
\ee
for $\{ r,s \} \in {\mathbb N}$. However, depending on the circumstances associated with a particular physical system, the connection
between normally ordered moments and Wigner characteristic function is more appropriate to our needs,
\be
\lb{mom2}
\llg {\bf c}^{\dagger r}(t) {\bf c}^{s}(t) \rg = (-1)^{s} \left. \frac{\upartial^{r+s}}{\upartial \xi^{r} \upartial \xi^{\ast s}} \,
{\rm e}^{\half | \xi |^{2}} \Lambda_{\dab}^{({\rm c})}(\xi,\xi^{\ast};t) \right|_{\xi , \xi^{\ast} = 0}
\ee
where $\Lambda_{\dab}^{({\rm c})}(\xi,\xi^{\ast};t) \equiv \Tr \lbk \ro(t) {\bf D}_{{\rm c}}(\xi) \rbk$ describes the symmetric ordering
of the creation and annihilation operators, and ${\bf D}_{{\rm c}}(\xi) = \exp (\xi {\bf c}^{\dagger} - \xi^{\ast} {\bf c} )$ is the
displacement operator. Thus, the initial task for both situations consists basically in the calculation of characteristic functions
associated with a physical system of interest described by the density operator $\ro(t)$. Consequently, the exact analytical expression
of the normally ordered moments will depend on the derivatives of $\Lambda_{\ene}^{({\rm c})}(\xi,\xi^{\ast};t)$ or ${\rm e}^{\half |
\xi |^{2}} \Lambda_{\dab}^{({\rm c})}(\xi,\xi^{\ast};t)$ with respect to $\xi$ and $\xi^{\ast}$ at the point $\xi , \xi^{\ast} = 0$.

Now, let us consider a feasible physical system modelled by a two-level atom interacting nonresonantly with a single-mode cavity field,
and driven additionally by an external quantum field through one open side of the cavity (the proposed experimental apparatus is 
sketched in Figure 1). Within the dipole and rotating-wave approximations, the dynamics of the atom-cavity system is governed basically
by the Hamiltonian ${\bf H} = {\bf H}_{0} + {\bf V}$, where \cite{r40}
\brr
{\bf H}_{0} &=& \hbar \om {\bf S} + \half \hbar \om \sig_{{\rm z}} \; , \nn \\
{\bf V} &=& \half \hbar \delta \sig_{{\rm z}} + \hbar \kappa_{\ief} ( {\bf A}^{\dagger} \sig_{-} + {\bf A} \sig_{+} ) \; . \nn
\err
Here, ${\bf S} \equiv {\bf n}_{{\rm a}} + {\bf n}_{{\rm b}}$ is a conserved quantity (${\bf n}_{{\rm a}} = {\bf a}^{\dagger} {\bf a}$
and ${\bf n}_{{\rm b}} = {\bf b}^{\dagger} {\bf b}$ correspond to the photon-number operators of the cavity and driving fields), 
${\bf A} \equiv \eps_{{\rm a}} {\bf a} + \eps_{{\rm b}} {\bf b}$ represents a quasi-mode operator with $\eps_{{\rm a}({\rm b})} =
\kappa_{{\rm a}({\rm b})} / \kappa_{\ief}$ ($\kappa_{\ief}^{2} = \kappa_{{\rm a}}^{2} + \kappa_{{\rm b}}^{2}$ is an effective coupling
constant between the atom and an effective field described by the quasi-mode operator), and $\delta = \om_{0} - \om$ is the detuning
frequency between the atomic transition frequency $\om_{0}$ and the cavity field frequency $\om$ (in particular, we assume the resonance
condition between the cavity and driving fields). In addition, the atomic spin-flip operators $\sig_{\pm}$ and $\sig_{{\rm z}}$ are
defined as $\sig_{+} \equiv | e \rg \llg g |$, $\sig_{-} \equiv | g \rg \llg e |$ and $\sig_{{\rm z}} \equiv | e \rg \llg e | - | g \rg
\llg g |$ ($| g \rg$ and $| e \rg$ correspond to the ground and excited states of the atom), which obey the commutation relations 
$[ \sig_{{\rm z}},\sig_{\pm} ] = \pm 2 \sig_{\pm}$ and $[ \sig_{+},\sig_{-} ] = \sig_{{\rm z}}$. Once that $[ {\bf H}_{0},{\bf V} ] =
0$, we obtain, in the interaction picture, the Hamiltonian ${\bf H}_{{\rm int}} = {\bf V}$, and this fact allows us to describe the
system through the well-known nonresonant JCM Hamiltonian for an atom interacting with the quasi-mode ${\bf A}$, and whose coupling
constant is given by $\kappa_{\ief}$. Consequently, the unitary time-evolution operator is the usual nonresonant JCM time-evolution
operator, namely $\opu(t) = \exp ( - \im {\bf V} t / \hbar)$.
\begin{figure}[!t]
\centering
\begin{minipage}[b]{0.70\linewidth}
\includegraphics[width=0.4\textwidth, angle=-90]{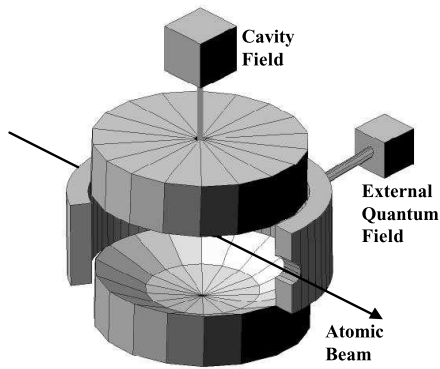}
\end{minipage}
\caption{Experimental scheme used in the description of the driven Jaynes-Cummings model. It consists of a high-$Q$ superconducting
cavity containing two electromagnetic fields prepared in the coherent states, and a beam of excited atoms interacting nonresonantly with
both the cavity field and the external quantum (driving) field.}
\end{figure}

After this brief introduction of the driven JCM, let us describe the density operator $\ro(t) = \opu(t) \ro(0) \opu^{\dagger}(t)$,
$\ro(0)$ being the density operator of the system at time $t=0$. For this purpose, we assume the atom is initially prepared in the
excited state and the cavity (a) and driving (b) fields are in the diagonal representation of coherent states -- i.e., the initial
density operator is written as $\ro(0) = \ro_{{\rm at}}(0) \otimes \ro_{{\rm ab}}(0)$, with $\ro_{{\rm at}}(0) = | e \rg \llg e |$ and
\bd
\ro_{{\rm ab}}(0) = \int \!\!\!\!\! \int \frac{d^{2} \alf_{{\rm a}} d^{2} \alf_{{\rm b}}}{\pi^{2}} \, P_{{\rm a}}(\alf_{{\rm a}})
P_{{\rm b}}(\alf_{{\rm b}}) | \alf_{{\rm a}},\alf_{{\rm b}} \rg \llg \alf_{{\rm a}},\alf_{{\rm b}} | \; ,
\ed
where $P(\alf)$ represents the Glauber-Sudarshan quasiprobability distribution function for each field. Thus, the matrix elements
$\ro_{{\rm ij}}(t)$ of the tripartite system in the atomic basis can be evaluated through the mathematical expressions
\brr
\ro_{11}(t) &=& \opu_{11}(t) \ro_{{\rm ab}}(0) \opu_{11}^{\dagger}(t) \; , \nn \\
\ro_{12}(t) &=& \opu_{11}(t) \ro_{{\rm ab}}(0) \opu_{21}^{\dagger}(t) \; , \nn \\
\ro_{21}(t) &=& \opu_{21}(t) \ro_{{\rm ab}}(0) \opu_{11}^{\dagger}(t) \; , \nn \\
\ro_{22}(t) &=& \opu_{21}(t) \ro_{{\rm ab}}(0) \opu_{21}^{\dagger}(t) \; . \nn 
\err
In these equalities, the matrix elements $\opu_{{\rm ij}}(t)$ are given by \cite{r41}
\brr
\opu_{11}(t) &=& \cos \lpar t \sqrt{\opb_{\ipa}} \rpar - \im \, \frac{\delta}{2} \frac{\sin \lpar t \sqrt{\opb_{\ipa}} \, \rpar}
{\sqrt{\opb_{\ipa}}} \; , \nn \\
\opu_{12}(t) &=& - \im \, \kappa_{\ief} \, \frac{\sin \lpar t \sqrt{\opb_{\ipa}} \, \rpar}{\sqrt{\opb_{\ipa}}} {\bf A} \; , \nn \\
\opu_{21}(t) &=& - \im \, \kappa_{\ief} \, {\bf A}^{\dagger} \, \frac{\sin \lpar t \sqrt{\opb_{\ipa}} \, \rpar}{\sqrt{\opb_{\ipa}}} 
\; , \nn \\
\opu_{22}(t) &=& \cos \lpar t \sqrt{\opf_{\ipa}} \rpar + \im \, \frac{\delta}{2} \frac{\sin \lpar t \sqrt{\opf_{\ipa}} \, \rpar}
{\sqrt{\opf_{\ipa}}} \; , \nn 
\err
with $\opf_{\ipa} = \kappa_{\ief}^{2} {\bf N}_{\ipa} + (\delta /2)^{2} {\bf 1}$, $\opb_{\ipa} = \opf_{\ipa} + \kappa_{\ief}^{2} 
{\bf 1}$, and ${\bf N}_{\ipa} = {\bf A}^{\dagger} {\bf A}$. Note that $\ro(t)$ describes the exact solution of the Schr\"{o}dinger
equation in the interaction picture with the nonresonant driven-JCM Hamiltonian. Using this solution we can establish exact analytical
expressions for the time evolution of some quantities (e.g., the Wigner characteristic function and the normally ordered moments for
each field) which allow us to shed more light on the quantitative aspects of entanglement in the system under consideration.

The Wigner characteristic functions for both the cavity and driving fields can be formally expressed in the integral form, with their
integrands having a common term that describes the product of the Glauber-Sudarshan quasiprobability distribution functions for each
field, and a kernel responsible for the entanglement among the constituents of the system -- namely
\be
\lb{e1}
\Lambda_{\dab}^{({\rm c})}(\xi,\xi^{\ast};t) = \int \!\!\!\!\! \int \frac{d^{2} \alf_{{\rm a}} d^{2} \alf_{{\rm b}}}{\pi^{2}} \,
P_{{\rm a}}(\alf_{{\rm a}}) P_{{\rm b}}(\alf_{{\rm b}}) \widetilde{{\rm K}}_{\xi,\xi^{\ast}}^{({\rm c})}(\alf_{{\rm a}},
\alf_{{\rm b}};t) \; ,
\ee
where
\be
\lb{e2}
\widetilde{{\rm K}}_{\xi,\xi^{\ast}}^{({\rm c})}(\alf_{{\rm a}},\alf_{{\rm b}};t) = \, _{11}^{({\rm c})}{\rm D}_{\xi,\xi^{\ast}}^{11}
(\alf_{{\rm a}},\alf_{{\rm b}};t) + \, _{21}^{({\rm c})}{\rm D}_{\xi,\xi^{\ast}}^{21} (\alf_{{\rm a}},\alf_{{\rm b}};t)
\ee
and
\be
\lb{e3}
_{{\rm ij}}^{({\rm c})}{\rm D}_{\xi,\xi^{\ast}}^{{\rm ij}}(\alf_{{\rm a}},\alf_{{\rm b}};t) = \llg \alf_{{\rm a}},\alf_{{\rm b}} 
| \opu_{{\rm ij}}^{\dagger}(t) {\bf D}_{{\rm c}}(\xi) \opu_{{\rm ij}}(t) | \alf_{{\rm a}},\alf_{{\rm b}} \rg \; .
\ee
The exact analytical expressions of the normally ordered moments for any kind of cavity and driving fields are then properly obtained
by means of Eq. (\ref{mom2}) when c = a,b. Henceforth the initial states of the cavity and driving fields will be fixed in the coherent
states $\{ | \nu_{{\rm a}({\rm b})} \rg \}$ throughout this paper. Such mathematical procedure diminishes considerably the technical
difficulties found in the double integrals over the complex $\alf_{{\rm a}}$- and $\alf_{{\rm b}}$-planes, because (i) the
Glauber-Sudarshan quasiprobability distribution functions for each field coincide with $P_{{\rm a}}(\alf_{{\rm a}}) = \pi \delta^{(2)}
(\alf_{{\rm a}}-\nu_{{\rm a}} )$ and $P_{{\rm b}}(\alf_{{\rm b}}) = \pi \delta^{(2)}( \alf_{{\rm b}}-\nu_{{\rm b}})$ ($\delta^{(2)}(z)$
denotes the two-dimensional delta function) and consequently, (ii) the Wigner characteristic function $\Lambda_{\dab}^{({\rm a},
{\rm b})}(\xi,\xi^{\ast};t)$ is equivalent to $\widetilde{{\rm K}}_{\xi,\xi^{\ast}}^{({\rm a},{\rm b})}(\nu_{{\rm a}},\nu_{{\rm b}};t)$.
Therefore, the normally ordered moments are expressed within this context as follows:
\brr
\lb{mom3}
\llg {\bf a}^{\dagger r}(t) {\bf a}^{s}(t) \rg &=& (-1)^{s} \left. \frac{\upartial^{r+s}}{\upartial \xi^{r} \upartial \xi^{\ast s}} \,
{\rm e}^{\half | \xi |^{2}} \, \widetilde{{\rm K}}_{\xi,\xi^{\ast}}^{({\rm a})}(\nu_{{\rm a}},\nu_{{\rm b}};t) 
\right|_{\xi,\xi^{\ast} = 0} \; , \\
\lb{mom4}
\llg {\bf b}^{\dagger r}(t) {\bf b}^{s}(t) \rg &=& (-1)^{s} \left. \frac{\upartial^{r+s}}{\upartial \xi^{r} \upartial \xi^{\ast s}} \,
{\rm e}^{\half | \xi |^{2}} \, \widetilde{{\rm K}}_{\xi,\xi^{\ast}}^{({\rm b})}(\nu_{{\rm a}},\nu_{{\rm b}};t) 
\right|_{\xi,\xi^{\ast} = 0} \; .
\err
The next task will consist basically in the calculations of the mean values (\ref{e3}) for the cavity and driving fields, and in the
derivation of theirs respective normally ordered moments.

\subsection{Cavity field} 

\hp The mathematical procedure adopted in the derivation process of the mean values (\ref{e3}) for the cavity field is based upon the
well-established results for $\mathfrak{su}(2)$ Lie algebra and its respective decomposition formulas \cite{r42}. After lengthy
calculations, the analytical expressions for the mean values assume the exact forms \cite{r40} 
\brr
{_{11}^{({\rm a})}}{\rm D}_{\xi,\xi^{\ast}}^{11}(\nu_{{\rm a}},\nu_{{\rm b}};t) &=& \sum_{m,m^{\prime} = 0}^{\infty} 
{_{11}}{\rm C}_{m,m^{\prime}}^{11}(\nu_{{\rm a}},\nu_{{\rm b}}) \, {\rm Y}_{\xi,\xi^{\ast}}^{(m,m^{\prime})}(\nu_{{\rm a}},
\nu_{{\rm b}}) \, F_{m}(t) F_{m^{\prime}}^{\ast}(t) \; , \nn \\
{_{21}^{({\rm a})}}{\rm D}_{\xi,\xi^{\ast}}^{21}(\nu_{{\rm a}},\nu_{{\rm b}};t) &=& \sum_{m,m^{\prime} = 0}^{\infty} 
{_{21}}{\rm C}_{m+1,m^{\prime}+1}^{21}(\nu_{{\rm a}},\nu_{{\rm b}}) \, {\rm Y}_{\xi,\xi^{\ast}}^{(m+1,m^{\prime}+1)}(\nu_{{\rm a}},
\nu_{{\rm b}}) \sqrt{(m+1)(m^{\prime}+1)} \, G_{m}(t) G_{m^{\prime}}^{\ast}(t) \; , \nn
\err
where the complex functions $_{{\rm ij}}{\rm C}_{m,m^{\prime}}^{{\rm ij}}(\nu_{{\rm a}},\nu_{{\rm b}})$ and 
${\rm Y}_{\xi,\xi^{\ast}}^{(m,m^{\prime})}(\nu_{{\rm a}},\nu_{{\rm b}})$ are given by
\bd
_{{\rm ij}}{\rm C}_{m,m^{\prime}}^{{\rm ij}}(\nu_{{\rm a}},\nu_{{\rm b}}) = \frac{1}{m^{\prime}!} \exp \lpar - \left| \eps_{{\rm a}}
\nu_{{\rm a}} + \eps_{{\rm b}} \nu_{{\rm b}} \right|^{2} \rpar \left| \eps_{{\rm a}} \nu_{{\rm a}} + \eps_{{\rm b}} \nu_{{\rm b}}
\right|^{2(j-i+m)} 
\ed
and
\bd
{\rm Y}_{\xi,\xi^{\ast}}^{(m,m^{\prime})}(\nu_{{\rm a}},\nu_{{\rm b}}) = \exp \lbr - \half | \xi |^{2} + 2 \im \ima \lbk 
\eps_{{\rm b}} \lpar \eps_{{\rm b}} \nu_{{\rm a}} - \eps_{{\rm a}} \nu_{{\rm b}} \rpar^{\ast} \xi \rbk \rbr \lbk \eps_{{\rm a}} \lpar
\eps_{{\rm a}} \nu_{{\rm a}} + \eps_{{\rm b}} \nu_{{\rm b}} \rpar^{\ast} \xi \rbk^{m^{\prime}-m} L_{m}^{(m^{\prime}-m)}\lpar 
\eps_{{\rm a}}^{2} | \xi |^{2} \rpar \; .
\ed
In these expressions, $L_{n}^{(k)}(z)$ denotes the associated Laguerre polynomials,
\bd
F_{m}(t) = \cos \lpar \frac{\Delta_{m} t}{2} \rpar - \im \, \frac{\delta}{\Delta_{m}} \sin \lpar \frac{\Delta_{m} t}{2} \rpar \qquad
\mbox{and} \qquad G_{m}(t) = - \im \, \frac{\Omega_{m}}{\Delta_{m}} \sin \lpar \frac{\Delta_{m} t}{2} \rpar
\ed
correspond to functions responsible for the time evolution of the mean values, with $\Omega_{m} = 2 \kappa_{\ief} \sqrt{m+1}$ being the effective Rabi frequency and $\Delta_{m}^{2} = \delta^{2} + \Omega_{m}^{2}$. 

Now, let us introduce the auxiliar function
\be
\lb{e4}
\mathfrak{A}_{m,m^{\prime}}^{(r,s)}(\nu_{{\rm a}},\nu_{{\rm b}}) = (-1)^{s} \left. \frac{\upartial^{r+s}}{\upartial \xi^{r} \upartial 
\xi^{\ast s}} \, {\rm e}^{\half | \xi |^{2}} {\rm Y}_{\xi,\xi^{\ast}}^{(m,m^{\prime})}(\nu_{{\rm a}},\nu_{{\rm b}}) 
\right|_{\xi , \xi^{\ast} = 0}
\ee
which has the following exact expression:
\bd
\mathfrak{A}_{m,m^{\prime}}^{(r,s)}(\nu_{{\rm a}},\nu_{{\rm b}}) = \left\{ 
\begin{array}{llll}
\displaystyle\frac{m^{\prime}!}{m!} \, \mathfrak{R}_{m,m^{\prime}}^{(r,s)}(\nu_{{\rm a}},\nu_{{\rm b}}) \,
\mathfrak{S}_{m,m^{\prime}}^{(s,r)} (\nu_{{\rm a}},\nu_{{\rm b}}) \qquad &(m^{\prime} \geq m)& \\ \\
\mathfrak{R}_{-m^{\prime},-m}^{(r,s)}(\nu_{{\rm a}},\nu_{{\rm b}}) \, \mathfrak{S}_{m^{\prime},m}^{(r,s)} (\nu_{{\rm a}},
\nu_{{\rm b}}) \, \mathfrak{T}^{(m-m^{\prime})}(\nu_{{\rm a}},\nu_{{\rm b}}) \qquad &(m^{\prime} \leq m)& 
\end{array} \right. 
\ed
with
\brr
\mathfrak{R}_{m,m^{\prime}}^{(r,s)}(\nu_{{\rm a}},\nu_{{\rm b}}) &=& \lbk \eps_{{\rm b}} \lpar \eps_{{\rm b}} \nu_{{\rm a}} - 
\eps_{{\rm a}} \nu_{{\rm b}} \rpar^{\ast} \rbk^{r} \lbk \eps_{{\rm b}} \lpar \eps_{{\rm b}} \nu_{{\rm a}} - \eps_{{\rm a}} 
\nu_{{\rm b}} \rpar \rbk^{s} \lbk \frac{\eps_{{\rm a}} \lpar \eps_{{\rm a}} \nu_{{\rm a}} + \eps_{{\rm b}} \nu_{{\rm b}}
\rpar^{\ast}}{\eps_{{\rm b}} \lpar \eps_{{\rm b}} \nu_{{\rm a}} - \eps_{{\rm a}} \nu_{{\rm b}} \rpar^{\ast}} \rbk^{m^{\prime}-m} 
\; , \nn \\
\mathfrak{S}_{m,m^{\prime}}^{(r,s)}(\nu_{{\rm a}},\nu_{{\rm b}}) &=& \sum_{k=0}^{\wp} k! \, L_{k}^{(m-k)}(0) \, L_{k}^{(r-k)}(0) \,
L_{m^{\prime}-m+k}^{(s-m^{\prime}+m-k)}(0) \, \mathfrak{T}^{(k)}(\nu_{{\rm a}},\nu_{{\rm b}}) \; , \nn \\
\mathfrak{T}^{(k)}(\nu_{{\rm a}},\nu_{{\rm b}}) &=& \lpar \frac{\eps_{{\rm a}}}{\eps_{{\rm b}} \left| \eps_{{\rm b}} \nu_{{\rm a}} -
\eps_{{\rm a}} \nu_{{\rm b}} \right|} \rpar^{2k} \; . \nn
\err
Here, $\wp \equiv \mbox{Min} [m,r,s-(m^{\prime}-m)]$ yields the smallest positive integer element of the set $\{m,r,s-(m^{\prime}-m)\}$.
One virtue of this auxiliar function is that it allows us to derive an exact analytical expression of the normally ordered moments 
for the cavity field, i.e.,
\brr
\lb{mom5}
\llg {\bf a}^{\dagger r}(t) {\bf a}^{s}(t) \rg_{\ell} &=& \sum_{m=0}^{\ell} \mathfrak{R}_{m,m}^{(r,s)}(\nu_{{\rm a}},\nu_{{\rm b}}) \,
{_{{\rm a}}}\Xi_{m,m}^{(r,s)} (\nu_{{\rm a}},\nu_{{\rm b}};t) \nn \\
& & + \sum_{m^{\prime}=0}^{\ell-1} \sum_{m=m^{\prime}+1}^{\ell} \mathfrak{R}_{-m^{\prime},-m}^{(r,s)} (\nu_{{\rm a}},\nu_{{\rm b}}) \,
\mathfrak{T}^{(m-m^{\prime})}(\nu_{{\rm a}},\nu_{{\rm b}}) \, {_{{\rm a}}}\Xi_{m,m^{\prime}}^{(r,s)} (\nu_{{\rm a}},\nu_{{\rm b}};t)
\nn \\
& & + \sum_{m=0}^{\ell-1} \sum_{m^{\prime}=m+1}^{\ell} \mathfrak{R}_{m,m^{\prime}}^{(r,s)}(\nu_{{\rm a}},\nu_{{\rm b}}) \,
{_{{\rm a}}}\Phi_{m,m^{\prime}}^{(s,r)}(\nu_{{\rm a}},\nu_{{\rm b}};t) \; ,
\err
where
\brr
{_{{\rm a}}}\Xi_{m,m^{\prime}}^{(r,s)}(\nu_{{\rm a}},\nu_{{\rm b}};t) &=& {_{11}}{\rm C}_{m,m^{\prime}}^{11}(\nu_{{\rm a}},
\nu_{{\rm b}}) \, \mathfrak{S}_{m^{\prime},m}^{(r,s)}(\nu_{{\rm a}},\nu_{{\rm b}}) \, F_{m}(t) F_{m^{\prime}}^{\ast}(t) \nn \\
& & + \sqrt{(m+1)(m^{\prime}+1)} \; {_{21}}{\rm C}_{m+1,m^{\prime}+1}^{21}(\nu_{{\rm a}},\nu_{{\rm b}}) \,
\mathfrak{S}_{m^{\prime}+1,m+1}^{(r,s)}(\nu_{{\rm a}},\nu_{{\rm b}}) \, G_{m}(t) G_{m^{\prime}}^{\ast}(t) , \nn \\
{_{{\rm a}}}\Phi_{m,m^{\prime}}^{(s,r)}(\nu_{{\rm a}},\nu_{{\rm b}};t) &=& {_{11}}{\rm C}_{m,m}^{11}(\nu_{{\rm a}},\nu_{{\rm b}}) \,
\mathfrak{S}_{m,m^{\prime}}^{(s,r)}(\nu_{{\rm a}},\nu_{{\rm b}}) \, F_{m}(t) F_{m^{\prime}}^{\ast}(t) \nn \\
& & + \sqrt{(m+1)(m^{\prime}+1)} \; {_{21}}{\rm C}_{m+1,m+1}^{21}(\nu_{{\rm a}},\nu_{{\rm b}}) \,
\mathfrak{S}_{m+1,m^{\prime}+1}^{(s,r)}(\nu_{{\rm a}},\nu_{{\rm b}}) \, G_{m}(t) G_{m^{\prime}}^{\ast}(t) . \nn 
\err
Note that the infinite sums present in ${_{{\rm ij}}^{({\rm a})}}{\rm D}_{\xi,\xi^{\ast}}^{{\rm ij}}(\nu_{{\rm a}},\nu_{{\rm b}};t)$
were substituted by finite sums, $\ell$ being the maximum value which guarantees the convergence of this expression (we have fixed
$\ell = 60$ in the numerical investigations). Indeed, exact expressions are reached only in the formal limit $\ell \rightarrow \infty$.
Following, we will establish similar results for the external driving field.

\subsection{External driving field}

\hp Adopting an analogous mathematical procedure for the external driving field, we verify that 
${_{{\rm ij}}^{({\rm b})}}{\rm D}_{\xi,\xi^{\ast}}^{{\rm ij}}(\nu_{{\rm a}},\nu_{{\rm b}};t)$ have similar structures to the previous
cases but differ in the dependence on the variables $\xi$ and $\xi^{\ast}$ -- i.e., the complex function 
${\rm Y}_{\xi,\xi^{\ast}}^{(m,m^{\prime})}(\nu_{{\rm a}},\nu_{{\rm b}})$ should be adequately replaced by 
\bd
{\rm Z}_{\xi,\xi^{\ast}}^{(m,m^{\prime})}(\nu_{{\rm a}},\nu_{{\rm b}}) = \exp \lbr - \half | \xi |^{2} + 2 \im \ima \lbk \eps_{{\rm a}}
\lpar \eps_{{\rm a}} \nu_{{\rm b}} - \eps_{{\rm b}} \nu_{{\rm a}} \rpar^{\ast} \xi \rbk \rbr \lbk \eps_{{\rm b}} \lpar \eps_{{\rm a}}
\nu_{{\rm a}} + \eps_{{\rm b}} \nu_{{\rm b}} \rpar^{\ast} \xi \rbk^{m^{\prime} - m} L_{m}^{(m^{\prime}-m)} \lpar \eps_{{\rm b}}^{2} 
| \xi |^{2} \rpar 
\ed
for each i,j = 1,2. Consequently, the auxiliar function 
\be
\lb{e5}
\mathfrak{B}_{m,m^{\prime}}^{(r,s)}(\nu_{{\rm a}},\nu_{{\rm b}}) = (-1)^{s} \left. \frac{\upartial^{r+s}}{\upartial \xi^{r} \upartial
\xi^{\ast s}} \, {\rm e}^{\half | \xi |^{2}} {\rm Z}_{\xi,\xi^{\ast}}^{(m,m^{\prime})}(\nu_{{\rm a}},\nu_{{\rm b}}) 
\right|_{\xi , \xi^{\ast} = 0}
\ee
has a central role in the present approach since its analytical expression
\bd
\mathfrak{B}_{m,m^{\prime}}^{(r,s)}(\nu_{{\rm a}},\nu_{{\rm b}}) = \left\{ 
\begin{array}{llll}
\displaystyle\frac{m^{\prime}!}{m!} \, \mathfrak{Z}_{m,m^{\prime}}^{(r,s)}(\nu_{{\rm a}},\nu_{{\rm b}}) \,
\mathfrak{W}_{m,m^{\prime}}^{(s,r)}(\nu_{{\rm a}},\nu_{{\rm b}}) \qquad &(m^{\prime} \geq m)& \\ \\ 
\mathfrak{Z}_{-m^{\prime},-m}^{(r,s)}(\nu_{{\rm a}},\nu_{{\rm b}}) \, \mathfrak{W}_{m^{\prime},m}^{(r,s)}(\nu_{{\rm a}},\nu_{{\rm b}})
\, \mathfrak{X}^{(m-m^{\prime})}(\nu_{{\rm a}},\nu_{{\rm b}}) \qquad &(m^{\prime} \leq m)&
\end{array} \right. 
\ed
with
\brr
\mathfrak{Z}_{m,m^{\prime}}^{(r,s)}(\nu_{{\rm a}},\nu_{{\rm b}}) &=& \lbk \eps_{{\rm a}} \lpar \eps_{{\rm a}} \nu_{{\rm b}} - 
\eps_{{\rm b}} \nu_{{\rm a}} \rpar^{\ast} \rbk^{r} \lbk \eps_{{\rm a}} \lpar \eps_{{\rm a}} \nu_{{\rm b}} - \eps_{{\rm b}} \nu_{{\rm a}}
\rpar \rbk^{s} \lbk \frac{\eps_{{\rm b}} \lpar \eps_{{\rm a}} \nu_{{\rm a}} + \eps_{{\rm b}} \nu_{{\rm b}} \rpar^{\ast}}{\eps_{{\rm a}}
\lpar \eps_{{\rm a}} \nu_{{\rm b}} - \eps_{{\rm b}} \nu_{{\rm a}} \rpar^{\ast}} \rbk^{m^{\prime}-m} \; , \nn \\
\mathfrak{W}_{m,m^{\prime}}^{(r,s)}(\nu_{{\rm a}},\nu_{{\rm b}}) &=& \sum_{k=0}^{\wp} k! \, L_{k}^{(m-k)}(0) \, L_{k}^{(r-k)}(0) \,
L_{m^{\prime}-m+k}^{(s-m^{\prime}+m-k)}(0) \, \mathfrak{X}^{(k)}(\nu_{{\rm a}},\nu_{{\rm b}}) \; , \nn \\
\mathfrak{X}^{(k)}(\nu_{{\rm a}},\nu_{{\rm b}}) &=& \lpar \frac{\eps_{{\rm b}}}{\eps_{{\rm a}} \left| \eps_{{\rm a}} \nu_{{\rm b}} -
\eps_{{\rm b}} \nu_{{\rm a}} \right|} \rpar^{2k} \; , \nn
\err
allows us to derive exact results for $\llg {\bf b}^{\dagger r}(t) {\bf b}^{s}(t) \rg$ through the Wigner characteristic function
$\Lambda_{\dab}^{({\rm b})}(\xi,\xi^{\ast};t)$. It is worth mentioning that the validity of this auxiliar function obeys the same
conditions verified for Eq. (\ref{e4}). 

Now, after some tedious calculations, the exact analytical expression of the normally ordered moments for the driving field can be
expressed as follows:
\brr
\lb{mom6}
\llg {\bf b}^{\dagger r}(t) {\bf b}^{s}(t) \rg_{\ell} &=& \sum_{m=0}^{\ell} \mathfrak{Z}_{m,m}^{(r,s)}(\nu_{{\rm a}},\nu_{{\rm b}}) \,
{_{{\rm b}}}\Xi_{m,m}^{(r,s)}(\nu_{{\rm a}},\nu_{{\rm b}};t) \nn \\
& & + \sum_{m^{\prime}=0}^{\ell-1} \sum_{m=m^{\prime}+1}^{\ell} \mathfrak{Z}_{-m^{\prime},-m}^{(r,s)}(\nu_{{\rm a}},\nu_{{\rm b}}) \,
\mathfrak{X}^{(m-m^{\prime})}(\nu_{{\rm a}},\nu_{{\rm b}}) \, {_{{\rm b}}}\Xi_{m,m^{\prime}}^{(r,s)}(\nu_{{\rm a}},\nu_{{\rm b}};t) 
\nn \\
& & + \sum_{m=0}^{\ell-1} \sum_{m^{\prime}=m+1}^{\ell} \mathfrak{Z}_{m,m^{\prime}}^{(r,s)}(\nu_{{\rm a}},\nu_{{\rm b}}) \,
{_{{\rm b}}}\Phi_{m,m^{\prime}}^{(s,r)}(\nu_{{\rm a}},\nu_{{\rm b}};t) \; , 
\err
where
\brr
{_{{\rm b}}}\Xi_{m,m^{\prime}}^{(r,s)}(\nu_{{\rm a}},\nu_{{\rm b}};t) &=& {_{11}}{\rm C}_{m,m^{\prime}}^{11}(\nu_{{\rm a}},
\nu_{{\rm b}}) \, \mathfrak{W}_{m^{\prime},m}^{(r,s)}(\nu_{{\rm a}},\nu_{{\rm b}}) \, F_{m}(t) F_{m^{\prime}}^{\ast}(t) \nn \\
& & + \sqrt{(m+1)(m^{\prime}+1)} \; {_{21}}{\rm C}_{m+1,m^{\prime}+1}^{21}(\nu_{{\rm a}},\nu_{{\rm b}}) \,
\mathfrak{W}_{m^{\prime}+1,m+1}^{(r,s)}(\nu_{{\rm a}},\nu_{{\rm b}}) \, G_{m}(t) G_{m^{\prime}}^{\ast}(t) , \nn \\
{_{{\rm b}}}\Phi_{m,m^{\prime}}^{(s,r)}(\nu_{{\rm a}},\nu_{{\rm b}};t) &=& {_{11}}{\rm C}_{m,m}^{11}(\nu_{{\rm a}},\nu_{{\rm b}}) \,
\mathfrak{W}_{m,m^{\prime}}^{(s,r)}(\nu_{{\rm a}},\nu_{{\rm b}}) \, F_{m}(t) F_{m^{\prime}}^{\ast}(t) \nn \\
& & + \sqrt{(m+1)(m^{\prime}+1)} \; {_{21}}{\rm C}_{m+1,m+1}^{21}(\nu_{{\rm a}},\nu_{{\rm b}}) \,
\mathfrak{W}_{m+1,m^{\prime}+1}^{(s,r)}(\nu_{{\rm a}},\nu_{{\rm b}}) \, G_{m}(t) G_{m^{\prime}}^{\ast}(t) . \nn 
\err
In the next section we will discuss, in principle, three specific applications for the normally ordered moments with emphasis on: (i)
the squeezing effect of the electromagnetic fields involved in this model; (ii) the nonlocal correlation measure which is based upon
the total variance of a pair of EPR type operators; and finally, (iii) the Shchukin-Vogel criterion.

\section{Applications}

\hp Recently, Shchukin and Vogel \cite{r16} have further developed the concept of the complete characterization of single-mode 
nonclassicality (basically, this concept is based on the negativity of the Glauber-Sudarshan quasiprobability distribution function and
requires an infinite hierarchy of conditions formulated either in terms of characteristic functions \cite{r17} or in terms of observable
moments \cite{r18}) with the aim of characterising the entanglement of bipartite quantum states for physical systems described by 
continuous variables. In fact, they have provided necessary and sufficient conditions for the partial transposition of bipartite 
harmonic quantum states to be nonnegative, such conditions being formulated as an infinite series of inequalities for the moments of the
states under investigation. Following the authors, `the violation of any inequality of this series is a sufficient condition for
entanglement.' In this section, we will show that the moments derived for the driven JCM not only are considered as a theoretical
realization of these conditions but also can be applied, for example, in the study of squeezing effect and in the numerical
investigation of some recent proposals of inseparability criteria. 

\subsection{Squeezing effect}

\hp Let us introduce two dimensionless quadrature operators ${\bf Q} = ({\bf c}^{\dagger} + {\bf c}) / \sqrt{2}$ and ${\bf P}= \im
({\bf c}^{\dagger} - {\bf c}) / \sqrt{2}$ for the electromagnetic field described by the q-number ${\bf c}$, satisfying the commutation
relation $[ {\bf Q},{\bf P} ] = \im {\bf 1}$. In general, the squeezing effect occurs when either the variance $\mathcal{V}_{{\bf Q}}(t)
\equiv \llg {\bf Q}^{2}(t) \rg - \llg {\bf Q}(t) \rg^{2}$ or $\mathcal{V}_{{\bf P}}(t) \equiv \llg {\bf P}^{2}(t) \rg - \llg {\bf P}(t)
\rg^{2}$ assumes values less than $1/2$ -- namely, $\mathcal{V}_{{\bf Q}}(t) < 1/2$ or $\mathcal{V}_{{\bf P}}(t) < 1/2$, but not both
simultaneously (for more details on the squeezing effect and its applications in physics, see the review article in \cite{r43}).

\begin{figure}[!t]
\centering
\begin{minipage}[b]{0.45\linewidth}
\includegraphics[width=5cm]{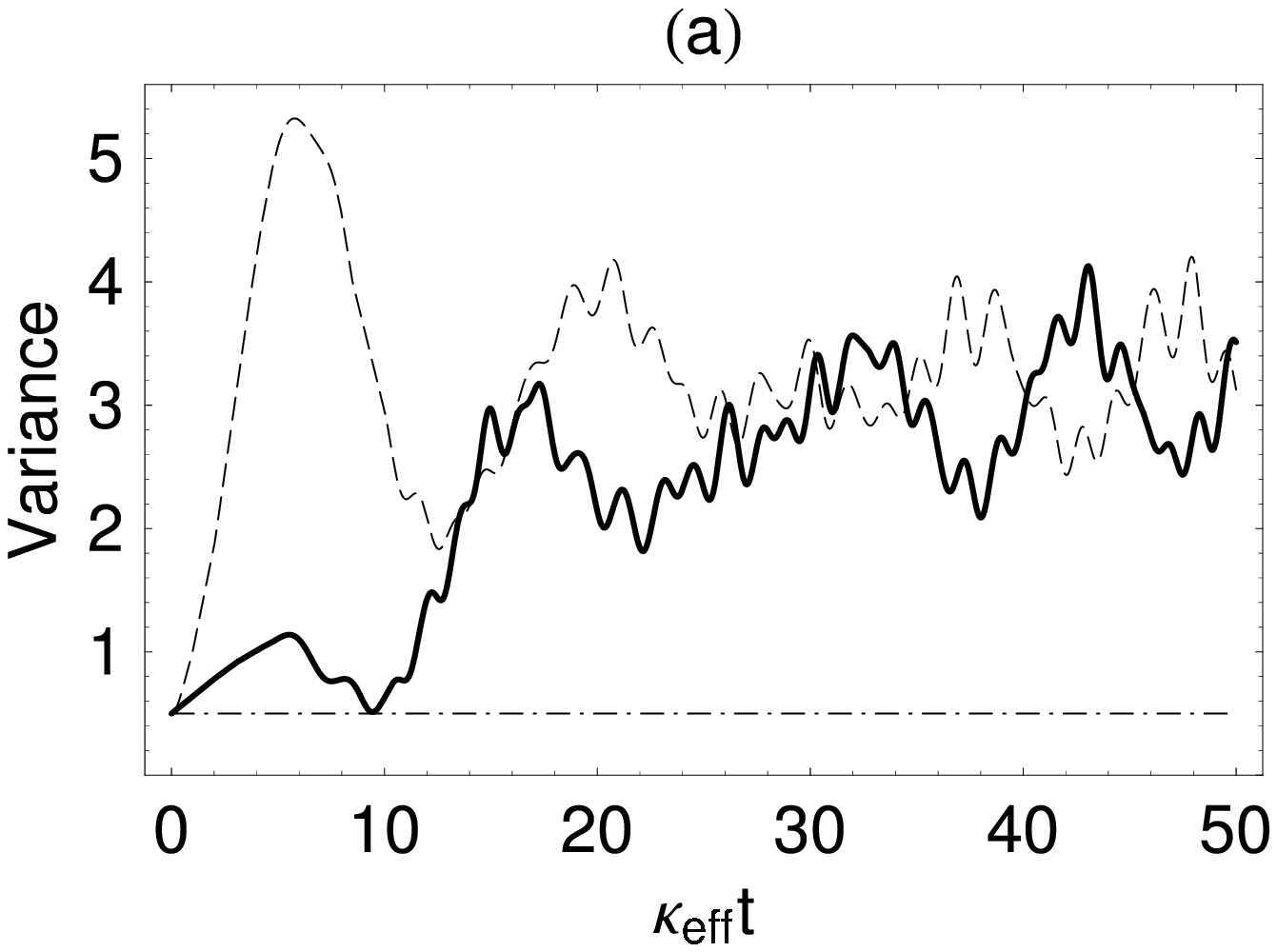}
\end{minipage} \hfill
\begin{minipage}[b]{0.45\linewidth}
\includegraphics[width=5cm]{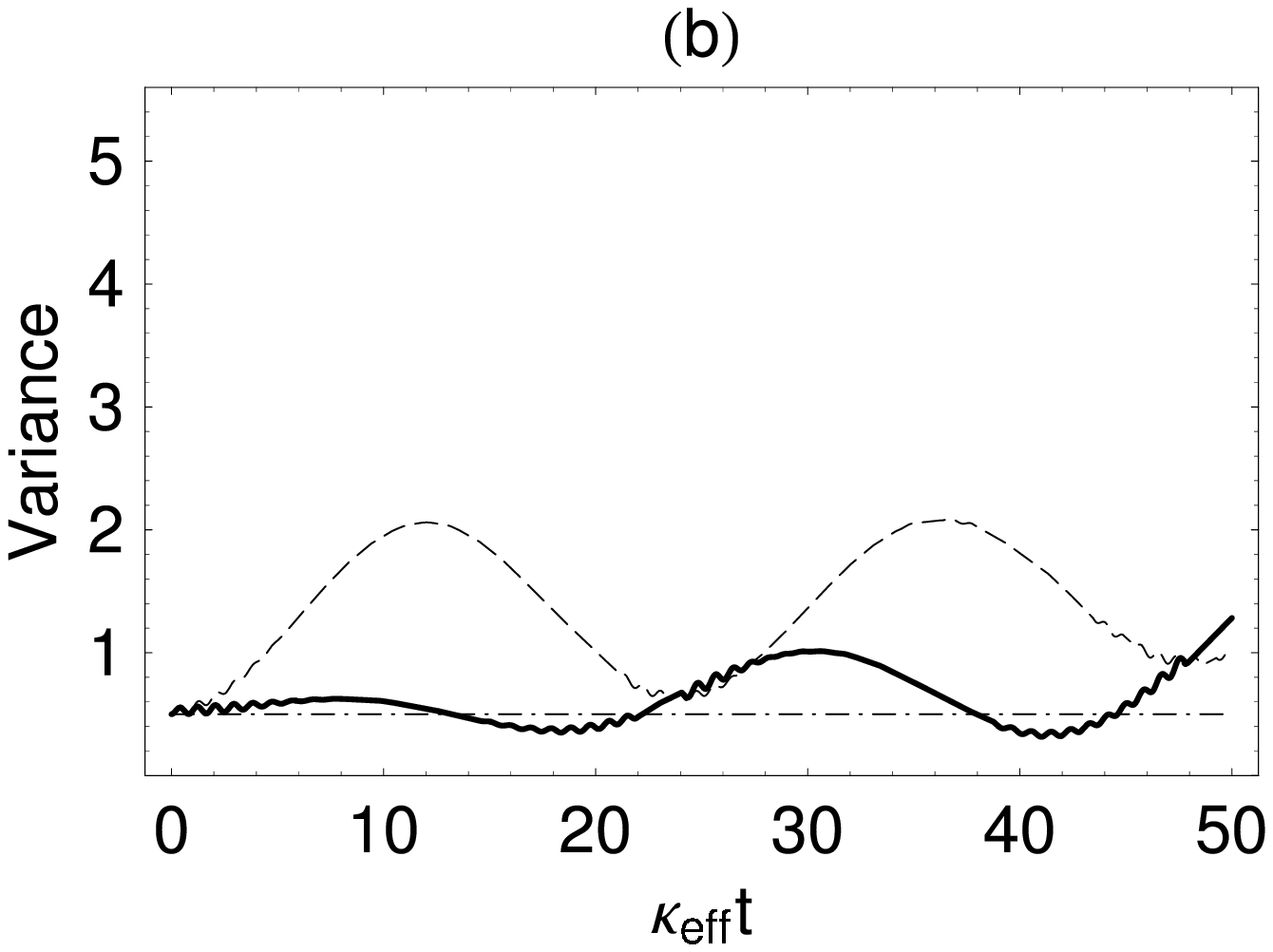}
\end{minipage} \hfill
\begin{minipage}[b]{0.45\linewidth}
\includegraphics[width=5cm]{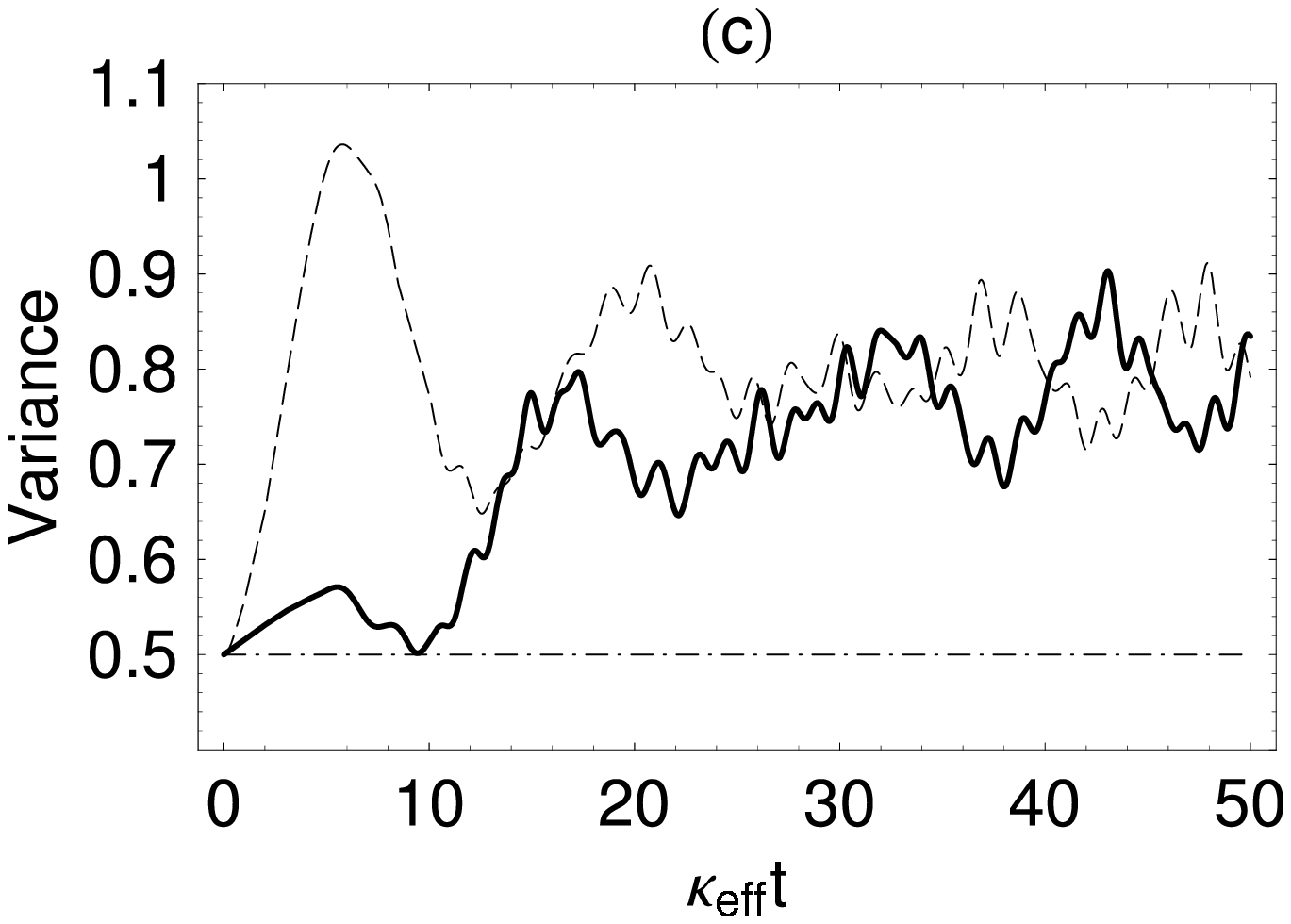}
\end{minipage} \hfill
\begin{minipage}[b]{0.45\linewidth}
\includegraphics[width=5cm]{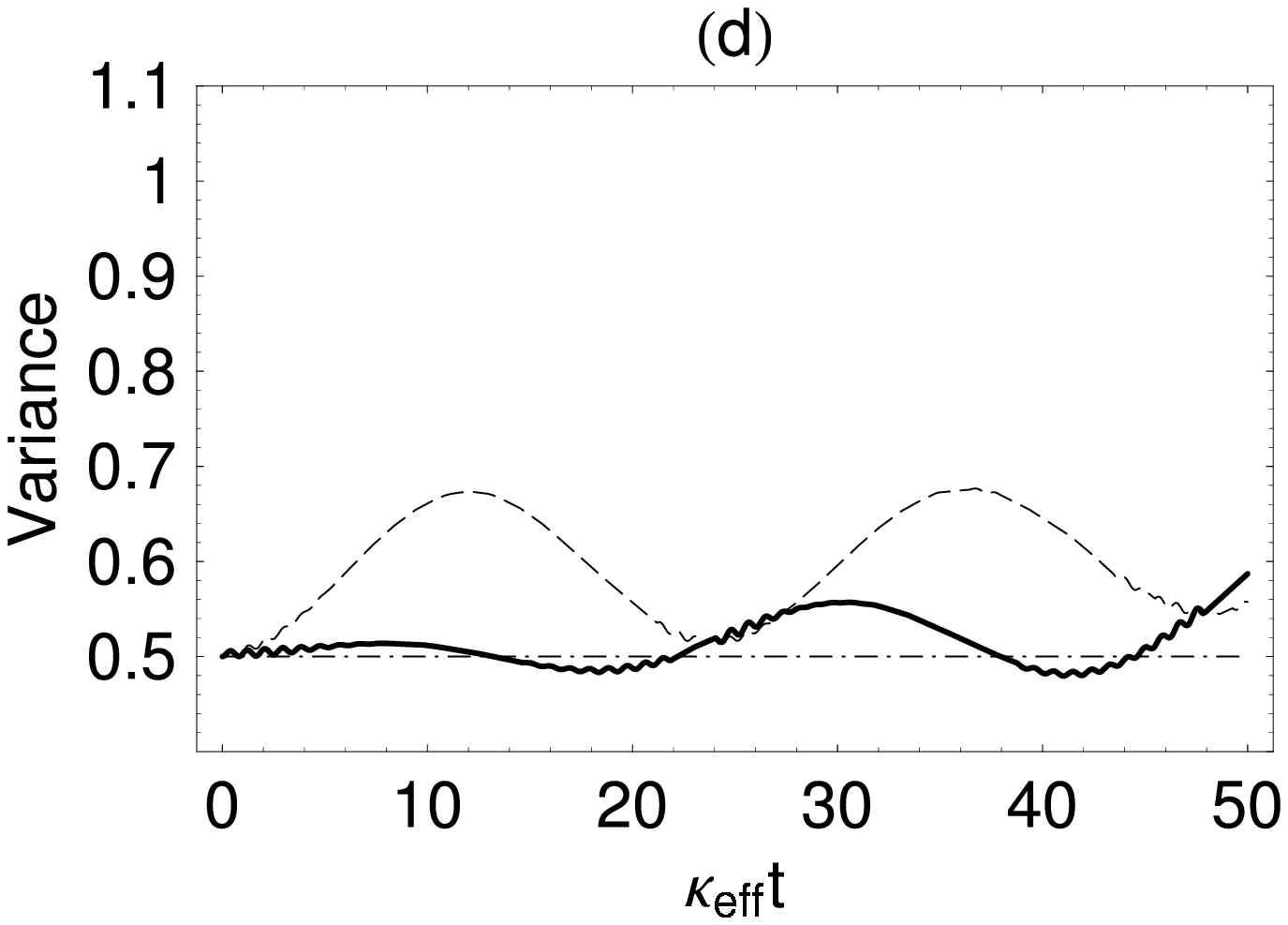}
\end{minipage}
\caption{Time evolution of variances associated with the quadrature operators ${\bf Q}_{{\rm a}({\rm b})}(t)$ and 
${\bf P}_{{\rm a}({\rm b})}(t)$ for the cavity (driving) field considering the atom initially prepared in the excited state, and both
the cavity and driving fields prepared in the coherent states. Pictures (a) and (b) represent the variances 
$\mathcal{V}_{{\bf Q}}^{({\rm a})}(t)$ (solid line) and $\mathcal{V}_{{\bf P}}^{({\rm a})}(t)$ (dashed line) for the following detuning
frequencies: (a) $\delta = 0$ (resonant) and (b) $\delta = 6 \kappa_{\ief}$ (nonresonant), with $\eps_{{\rm a}} = 3/ \sqrt{10}$,
$\eps_{{\rm b}} = 1/ \sqrt{10}$, $| \nu_{{\rm a}} | = 1$, and $| \nu_{{\rm b}} | = 2$ fixed. In addition, pictures (c) and (d) 
correspond to $\mathcal{V}_{{\bf Q}}^{({\rm b})}(t)$ (solid line) and $\mathcal{V}_{{\bf P}}^{({\rm b})}(t)$ (dashed line) for the same
set of parameters adopted in (a) and (b), respectively. The dot-dashed line present in all pictures refers to the variances of the
coherent states (also described in literature as minimum-uncertainty states). Note that the variances show squeezing effect, this effect
being connected with the amplitude $| \nu_{{\rm a}({\rm b})} |$ of the cavity (driving) field and the detuning frequency $\delta$.}
\end{figure}
Figure 2 shows the plots of $\mathcal{V}_{{\bf Q}}^{({\rm a},{\rm b})}(t)$ (solid line) and $\mathcal{V}_{{\bf P}}^{({\rm a},
{\rm b})}(t)$ (dashed line) versus $\kappa_{\ief}t \in [0,50]$ when the atom-field system is resonant (a),(c) $\delta = 0$ and 
nonresonant (b),(d) $\delta = 6 \kappa_{\ief}$, for $\eps_{{\rm a}} = 3/\sqrt{10}$, $\eps_{{\rm b}} = 1/ \sqrt{10}$, $| \nu_{{\rm a}} |
= 1$, and $| \nu_{{\rm b}} | = 2$ fixed. In particular, pictures (a) and (b) represent the variances associated with the quadrature
operators ${\bf Q}_{{\rm a}}(t)$ and ${\bf P}_{{\rm a}}(t)$ (cavity field), while (c) and (d) correspond to the operators 
${\bf Q}_{{\rm b}}(t)$ and ${\bf P}_{{\rm b}}(t)$ (driving field). A first analysis of these pictures leads us to observe that the
squeezing effect can be linked directly with the amplitude of the cavity (driving) field $| \nu_{{\rm a}({\rm b})} |$ and the detuning
frequency $\delta$ \cite{r23,r44}. Indeed, numerical investigations for $| \nu_{{\rm a}({\rm b})} | \gg 1$ and $\delta \geq 0$ show that
$\mathcal{V}_{{\bf Q}}^{({\rm a},{\rm b})}(t)$ and $\mathcal{V}_{{\bf P}}^{({\rm a},{\rm b})}(t)$ have the same asymptotic value $1/2$,
which minimizes the Heisenberg uncertainty relation $\mathcal{V}_{{\bf Q}}^{({\rm a},{\rm b})}(t) \mathcal{V}_{{\bf P}}^{({\rm a}, 
{\rm b})}(t) \geq 1/4$. Furthermore, an interesting aspect of this important effect refers to picture (d), where the variance
$\mathcal{V}_{{\bf Q}}^{({\rm b})}(t)$ associated with the single-mode external quantum field exhibits the squeezing effect for $\delta
\neq 0$ (from the experimental point of view, this theoretical prediction can be confirmed through balanced homodyne detections or
similar techniques -- for instance, see the experimental apparatus used by Josse and co-workers \cite{r14} in the demonstration of both
quadrature and polarization entanglement generated via the interaction between a coherent linearly polarized field and cold atoms in a
high finesse optical cavity). Beyond these important points on the squeezing effect, it is worth mentioning that the similarities
between the different patterns of curves observed in (a),(c) for $\delta = 0$ and (b),(d) when $\delta = 6 \kappa_{\ief}$, can be
explained by means of the entanglement dynamics verified in $\kappa_{\ief}t > 0$ among the constituent parts of the tripartite system
(in this case, the continuous variable entanglement for the cavity and driving fields is obtained by tracing over the atomic variables).
Thus, for the specific initial states of the cavity and driving fields adopted in this work, we can conclude that nonclassical effects
and entanglement present an important link in the driven JCM, and this fact leads us to investigate some recent proposals of 
inseparability criteria for continuous bipartite quantum states. Next, we will focus upon the following question: ``Can EPR uncertainty
be considered the first quantitative characterization of the entanglement properties of the system under consideration?"

\subsection{EPR uncertainty}

\hp For any bipartite physical system defined in a Hilbert space $\mathcal{H} = \mathcal{H}_{{\rm a}} \otimes \mathcal{H}_{{\rm b}}$
and described by an arbitrary normalized density operator $\ro(t)$, the total variance of a pair of EPR-like operators -- such as
$({\bf Q}_{{\rm a}} - {\bf Q}_{{\rm b}}) / \sqrt{2}$ and $({\bf P}_{{\rm a}} + {\bf P}_{{\rm b}}) / \sqrt{2}$, with 
${\bf Q}_{{\rm a}({\rm b})}$ and ${\bf P}_{{\rm a}({\rm b})}$ satisfying the commutation relation $[ {\bf Q}_{\alf},{\bf P}_{\bet} ] =
\im \delta_{\alf,\bet} {\bf 1}$ for $\alf,\bet = {\rm a},{\rm b}$ -- allows us to establish  the mathematical relation \cite{r6}
\be
\lb{e8}
{\rm I}_{{\rm a},{\rm b}}(t) = \half \lbk \mathcal{V}_{{\bf Q}}^{({\rm a})}(t) + \mathcal{V}_{{\bf Q}}^{({\rm b})}(t) +
\mathcal{V}_{{\bf P}}^{({\rm a})}(t) + \mathcal{V}_{{\bf P}}^{({\rm b})}(t) - 2 \, {\rm Cov}_{{\bf Q}}^{({\rm a},{\rm b})}(t) + 2 \,
{\rm Cov}_{{\bf P}}^{({\rm a},{\rm b})}(t) \rbk \geq 0 
\ee
where the covariance ${\rm Cov}_{{\bf X}}^{({\rm a},{\rm b})}(t)$ for the operators ${\bf X}_{{\rm a}} \in \mathcal{H}_{{\rm a}}$ and
${\bf X}_{{\rm b}} \in \mathcal{H}_{{\rm b}}$ is given by
\be
\lb{e9}
{\rm Cov}_{{\bf X}}^{({\rm a},{\rm b})}(t) \equiv \half \Tr \lbk \ro(t) \{ {\bf X}_{{\rm a}},{\bf X}_{{\rm b}} \} \rbk - \Tr \lbk \ro(t)
{\bf X}_{{\rm a}} \rbk \Tr \lbk \ro(t) {\bf X}_{{\rm b}} \rbk \; ,
\ee
and $\Tr \lbk \ro(t) \{ {\bf X}_{{\rm a}},{\bf X}_{{\rm b}} \} \rbk$ represents the anticommutation relation mean value. Note that 
${\rm I}_{{\rm a},{\rm b}}(t)$ reaches the value zero when $\ro(t)$ behaves as an idelized EPR type density operator. Furthermore, for
states with Gaussian statistics, ${\rm I}_{{\rm a},{\rm b}}(t) < 1$ not only implies in the existence of nonlocal correlations but also
establishes a sufficient condition for entanglement (in particular, this inseparabi\-lity criterion has already been used in several
experiments to demonstrate continuous variable entanglement -- for instance, see \cite{r14} and references therein). The additional
condition ${\rm I}_{{\rm a},{\rm b}}(t) \geq 1$ characterizes separable states into this context. Hence, the quantity ${\rm I}_{{\rm a},
{\rm b}}(t)$ measures the degree of nonlocal correlations and the inequality (\ref{e8}) defines the EPR uncertainty for $\ro(t)$. In
Appendix A we determine explicitly the generalized moments $\llg {\bf a}^{\dagger p} (t) {\bf a}^{q}(t){\bf b}^{\dagger r}(t) 
{\bf b}^{s}(t) \rg$, which are necessary in the numerical evaluation of the covariance function.

\begin{figure}[!t]
\centering
\begin{minipage}[b]{0.45\linewidth}
\includegraphics[width=5cm]{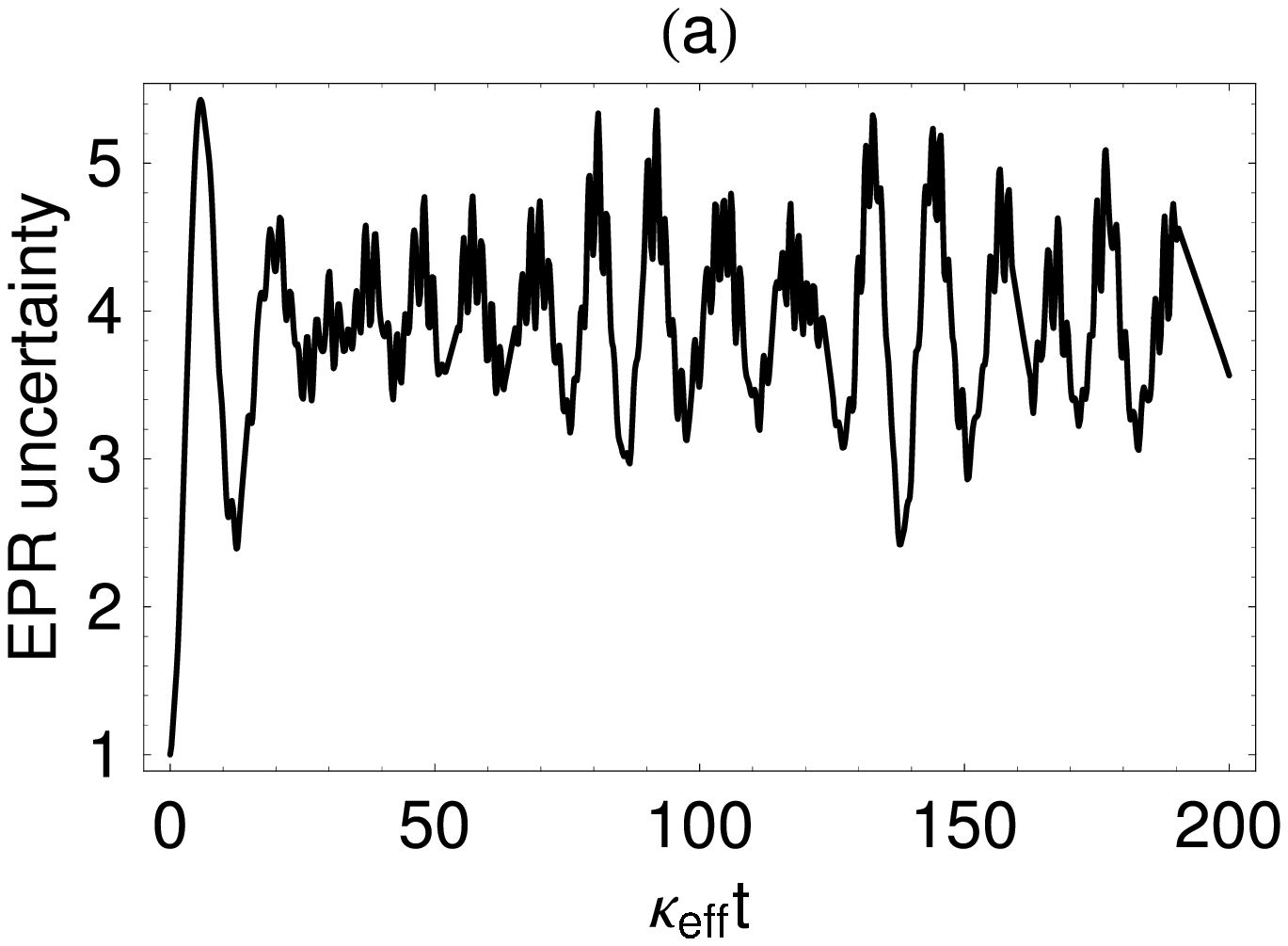}
\end{minipage} \hfill
\begin{minipage}[b]{0.45\linewidth}
\includegraphics[width=5cm]{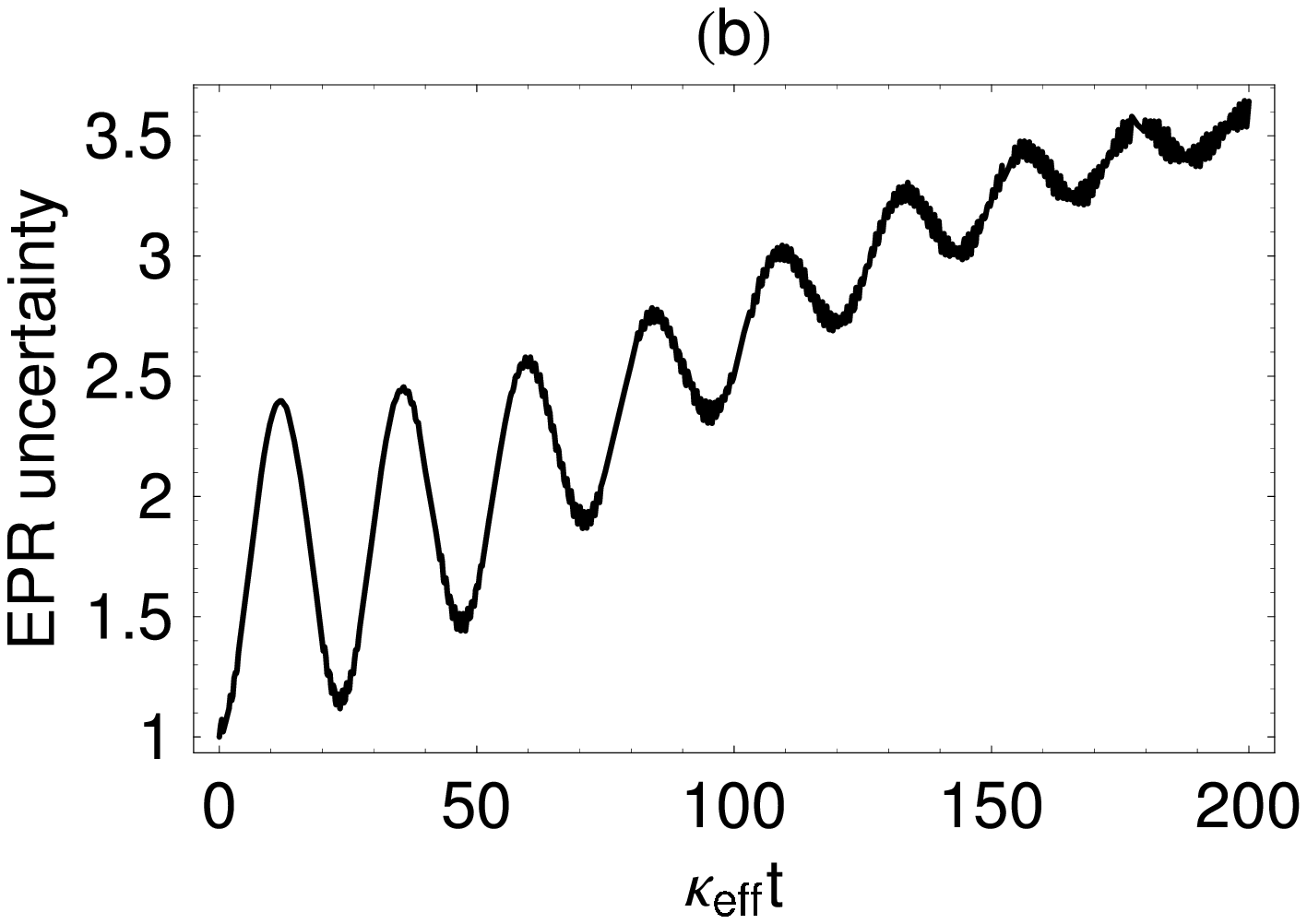}
\end{minipage} \hfill
\begin{minipage}[b]{0.45\linewidth}
\includegraphics[width=5cm]{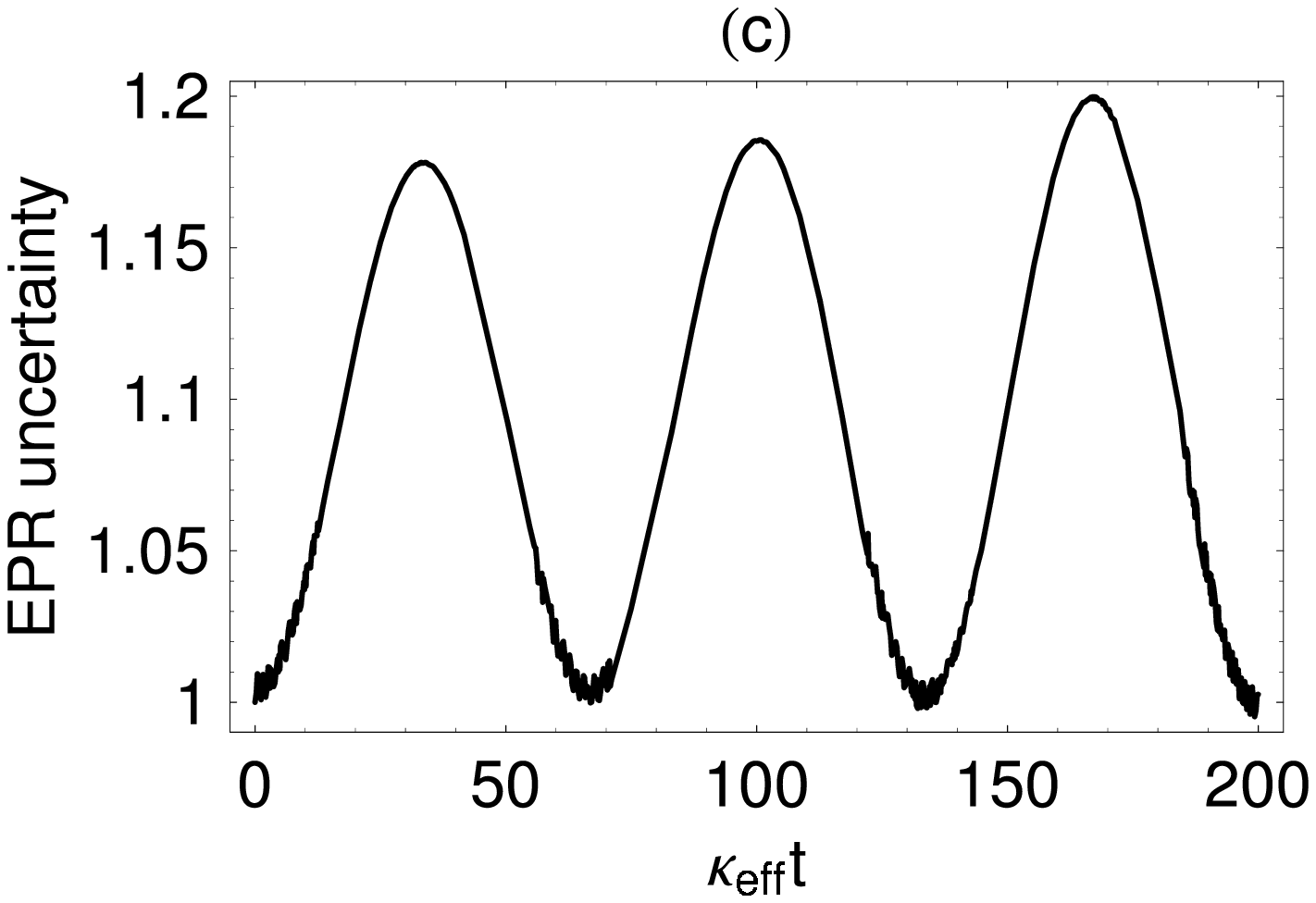}
\end{minipage} \hfill
\begin{minipage}[b]{0.45\linewidth}
\includegraphics[width=5cm]{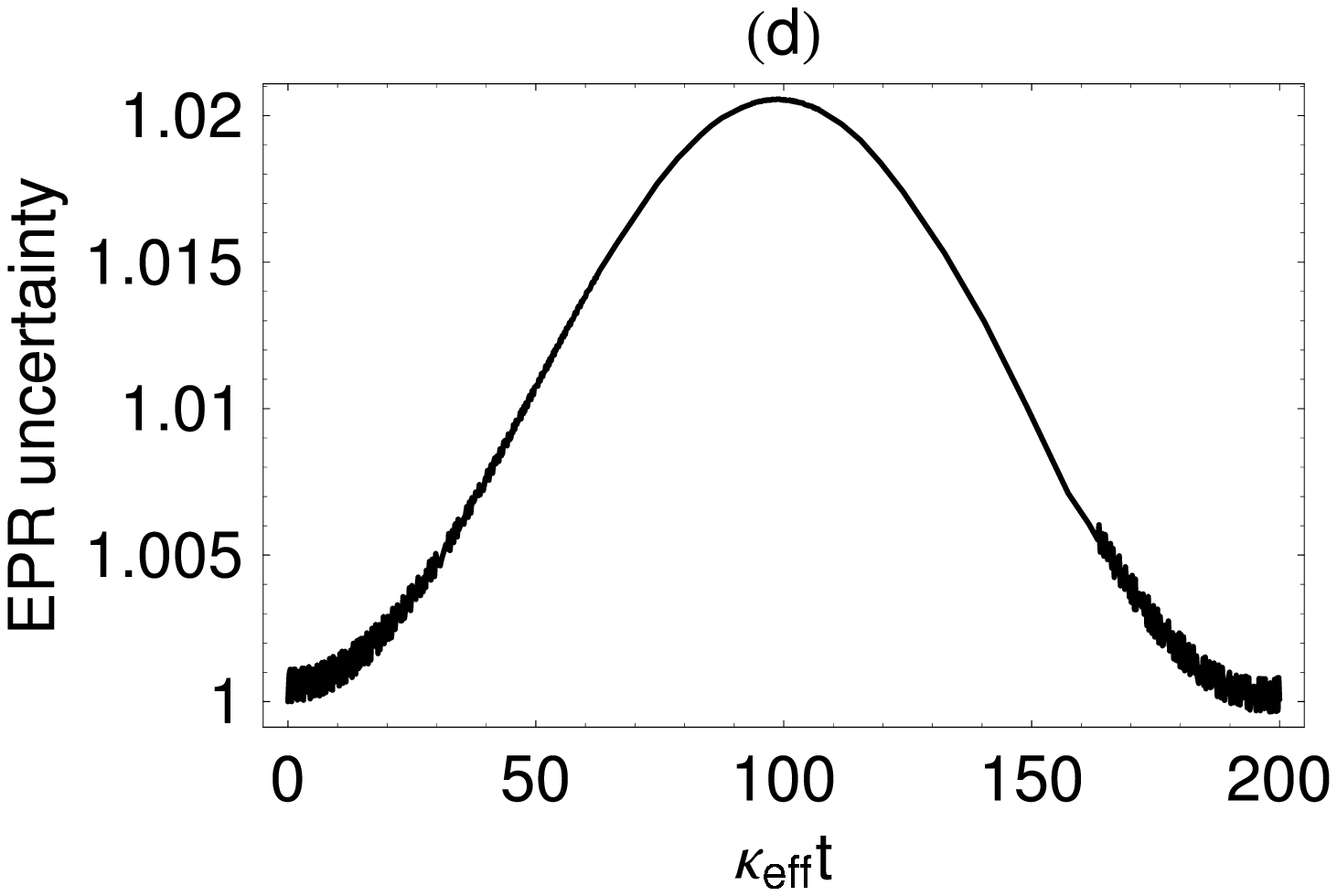}
\end{minipage}
\caption{Time evolution of ${\rm I}_{{\rm a},{\rm b}}(t)$ versus $\kappa_{\ief} t$ for $\eps_{{\rm a}} = 3/ \sqrt{10}$, 
$\eps_{{\rm b}} = 1/ \sqrt{10}$, $| \nu_{{\rm a}} | = 1$, and $| \nu_{{\rm b}} | = 2$ fixed, with different values of detuning
frequency: (a) $\delta = 0$ (resonant regime), (b) $\delta = 6 \kappa_{\ief}$, (c) $\delta = 20 \kappa_{\ief}$, and (d) 
$\delta = 60 \kappa_{\ief}$. Although the variances associated with the operators ${\bf Q}_{{\rm a}}(t)$ and ${\bf Q}_{{\rm b}}(t)$
have shown a nonclassical behaviour (squeezing effect) for the nonresonant regime, the EPR uncertainty can not be considered the first
quantitative characterization of the entanglement properties in the driven JCM since ${\rm I}_{{\rm a},{\rm b}}(t) \geq 1$ for all 
$t \geq 0$ (this case characterizes separable states). Beyond this important fact, it is worth mentioning that different initial
conditions for the cavity and driving fields can be considered in the study of squeezing and entanglement. However, additional
inseparability criteria must be employed in the quantitative characterization of the entanglement properties such as, for example, the
Shchukin-Vogel criterion \cite{r16}.}
\end{figure}
Figure 3 shows the plots of ${\rm I}_{{\rm a},{\rm b}}(t)$ versus $\kappa_{\ief}t$ for different values of detuning frequency: (a)
$\delta = 0$, (b) $\delta = 6 \kappa_{\ief}$, (c) $\delta = 20 \kappa_{\ief}$, and (d) $\delta = 60 \kappa_{\ief}$, where we have
adopted the same set of parameters established in the previous figure. Note that EPR uncertainty is completely satisfied in all
pictures, but unfortunately it {\sl does not} reveal entanglement in the driven JCM since ${\rm I}_{{\rm a},{\rm b}}(t) \geq 1$.
According to Giedke and co-workers \cite{r12}, the condition ${\rm I}_{{\rm a},{\rm b}}(t) < 1$ is met `only if at least one of the
uncertainties of $({\bf Q}_{{\rm a}}-{\bf Q}_{{\rm b}})/ \sqrt{2}$ or $({\bf P}_{{\rm a}} + {\bf P}_{{\rm b}})/ \sqrt{2}$ lies below
$1$ (the standard quantum limit). This implies that the corresponding states must possess a certain squeezing.' In this sense, despite
$\mathcal{V}_{{\bf Q}}^{({\rm a})}(t)$ and $\mathcal{V}_{{\bf Q}}^{({\rm b})}(t)$ have shown signatures of a nonclassical behaviour in
pictures 2(b) and 2(d) for $\delta = 6 \kappa_{\ief}$, the inseparability criterion does not demonstrate the link between squeezing and
entanglement in the theoretical model under study. Similar result was obtained by Shchukin and Vogel \cite{r16} for the example of an
entangled quantum composed of two coherent states, whose expression reads as
\bd
| \psi_{-} \rg = \lbk 2 \lpar 1 - {\rm e}^{-2 ( | \alf |^{2} + | \bet |^{2} )} \rpar \rbk^{-1/2} \lpar | \alf,\bet \rg - | 
-\alf,-\bet \rg \rpar \; .
\ed
In particular, the authors showed that the inseparability criteria established in \cite{r6,r7} fail to demonstrate the entanglement of
$| \psi_{-} \rg$. Consequently, our results lead us to investigate new inseparability criteria based in principle on certain series of
inequalities involving higher-order moments which can be applied to a variety of quantum states. Thus, the next step will consist in
the study of simple subdeterminants derived from the Shchukin-Vogel criterion \cite{r16} for entanglement.

\subsection{Shchukin-Vogel criterion}

\hp Recently, Shchukin and Vogel \cite{r16} derived a hierarchy of necessary and sufficient conditions for the negativity of the
partial transposition of bipartite quantum states, which are formulated in terms of observable moments associated with a variety of
quantum states. This hierarchy is basically characterized by a set of inseparability criteria (or sufficient conditions for 
entanglement) that generalizes some previous purposes established in the literature (e.g., see Refs.
\cite{r6,r7,r8,r9,r10,r11,r12,r13,r19}). For practical reasons, let us formulate the Shchukin-Vogel criterion as follows:
\begin{center}
\begin{minipage}[t]{13cm}
{\bf Shchukin-Vogel criterion.} The partial transposition of any bipartite quantum state is nonnegative if all the determinants derived
from the $N$th order determinant
\brr
\Gamma_{\ene} = \left|
\begin{array}{cccccc}
1 & \llg {\bf a} \rg & \llg {\bf a}^{\dagger} \rg & \llg {\bf b}^{\dagger} \rg & \llg {\bf b} \rg & \cdots \\
\llg {\bf a}^{\dagger} \rg & \llg {\bf a}^{\dagger} {\bf a} \rg & \llg {\bf a}^{\dagger 2} \rg & \llg {\bf a}^{\dagger} {\bf b}^{\dagger}
\rg & \llg {\bf a}^{\dagger} {\bf b} \rg & \cdots \\
\llg {\bf a} \rg & \llg {\bf a}^{2} \rg & \llg {\bf a} {\bf a}^{\dagger} \rg & \llg {\bf a} {\bf b}^{\dagger} \rg & \llg {\bf a} {\bf b}
\rg & \cdots \\
\llg {\bf b} \rg & \llg {\bf a} {\bf b} \rg & \llg {\bf a}^{\dagger} {\bf b} \rg & \llg {\bf b}^{\dagger} {\bf b} \rg & \llg {\bf b}^{2}
\rg & \cdots \\
\llg {\bf b}^{\dagger} \rg & \llg {\bf a} {\bf b}^{\dagger} \rg & \llg {\bf a}^{\dagger} {\bf b}^{\dagger} \rg & \llg {\bf b}^{\dagger 2}
\rg & \llg {\bf b} {\bf b}^{\dagger} \rg & \cdots \\
\vdots & \vdots & \vdots & \vdots & \vdots & \cdots \\
\end{array}
\right| \nn
\err
are nonnegative (in other words, $\forall \textrm{N}$ we obtain $\Gamma_{\ene} \geq 0$). Howe\-ver, if there exists a negative
determinant (namely, $\exists \textrm{N}$ such that $\Gamma_{\ene} < 0$) the negativity of the partial transposition has been
demonstrated -- this fact provides a sufficient condition for any bipartite quantum state under investigation to be entangled.
\end{minipage}
\end{center}
%
\begin{figure}[!t]
\centering
\begin{minipage}[b]{0.45\linewidth}
\includegraphics[width=5cm]{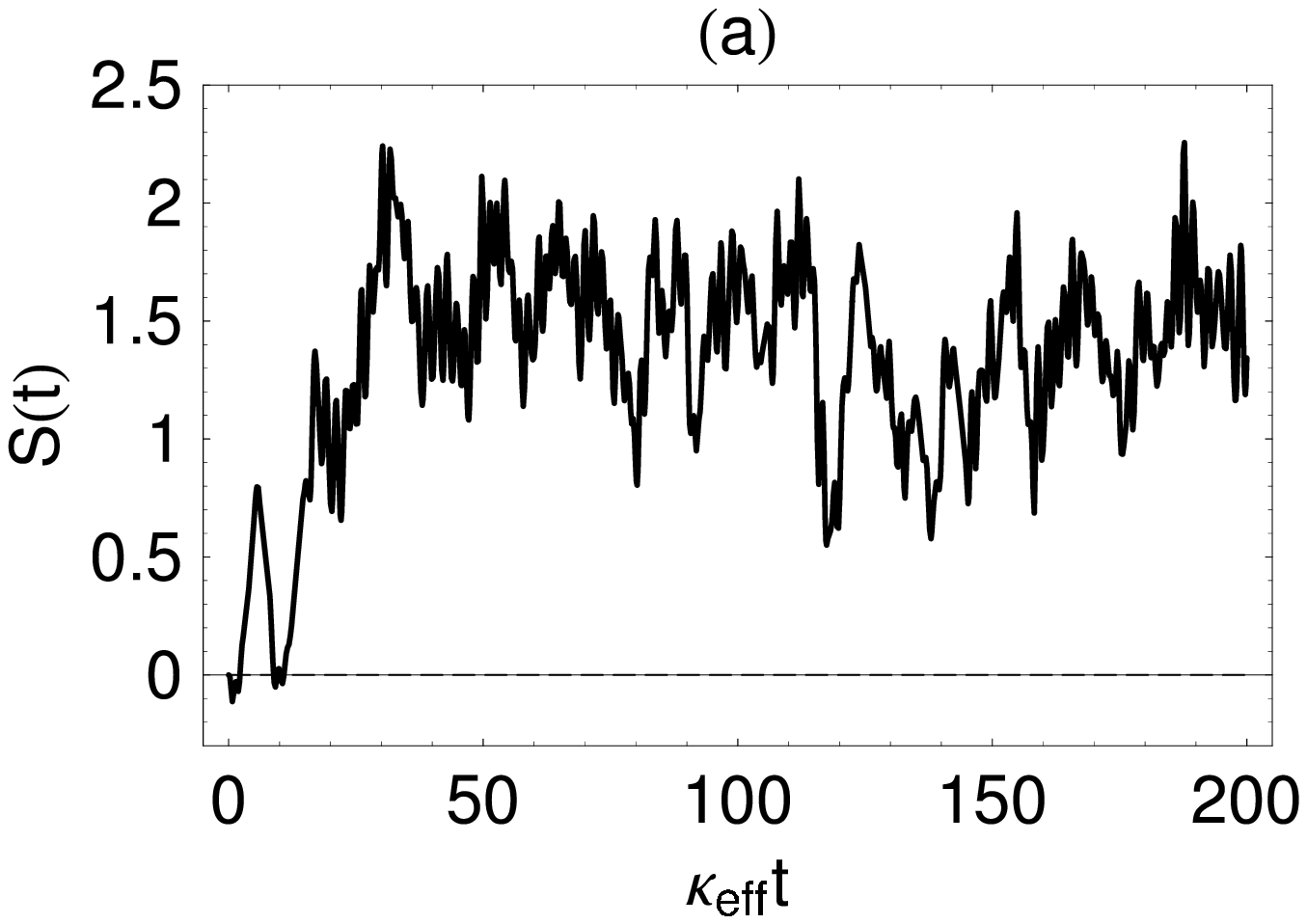}
\end{minipage} \hfill
\begin{minipage}[b]{0.45\linewidth}
\includegraphics[width=5cm]{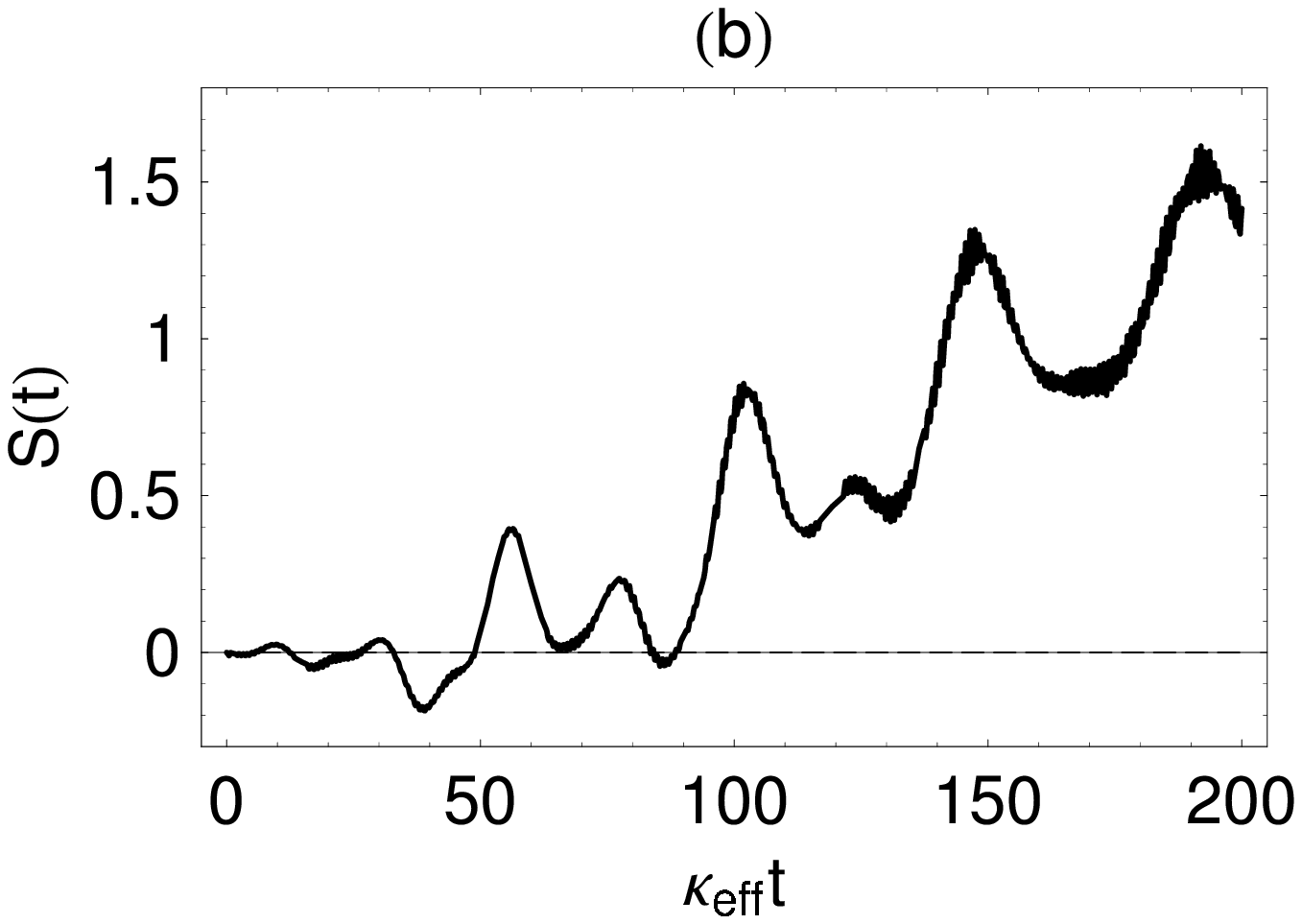}
\end{minipage}
\caption{Time evolution of $\textrm{S}(t)$ versus $\kappa_{\ief} t$ for different values of detuning frequency: (a) $\delta = 0$
(resonant regime) and (b) $\delta = 6 \kappa_{\ief}$ (nonresonant regime), where we have adopted the same set of parameters established
in Figure 2. Note that $\textrm{S}(t)$ assumes negative values in both pictures and this fact corroborates the Shchukin-Vogel criterion
for entanglement.}
\end{figure}

To illustrate this criterion within the driven JCM we consider, in principle, the time-dependent subdeterminant
\brr
\lb{e10}
\textrm{S}(t) = \left|
\begin{array}{ccc}
1 & \llg {\bf b}^{\dagger}(t) \rg & \llg {\bf a}(t) {\bf b}^{\dagger}(t) \rg \\
\llg {\bf b}(t) \rg & \llg {\bf b}^{\dagger}(t) {\bf b}(t) \rg & \llg {\bf a}(t) {\bf b}^{\dagger}(t) {\bf b}(t) \rg \\
\llg {\bf a}^{\dagger}(t) {\bf b}(t) \rg & \llg {\bf a}^{\dagger}(t) {\bf b}^{\dagger}(t) {\bf b}(t) \rg & \llg {\bf a}^{\dagger}(t) 
{\bf a}(t) {\bf b}^{\dagger}(t) {\bf b}(t) \rg \\
\end{array}
\right|
\err
where the mean values $\llg {\bf a}^{\dagger p}(t) {\bf a}^{q}(t) {\bf b}^{\dagger r}(t) {\bf b}^{s}(t) \rg$ can be evaluated through
the results obtained in Appendix A. Figure 4 shows the plots of $\textrm{S}(t)$ versus $\kappa_{\ief} t$ for $\eps_{{\rm a}}=3 /
\sqrt{10}$, $\eps_{{\rm b}}=1 /\sqrt{10}$, $| \nu_{{\rm a}} | = 1$, and $| \nu_{{\rm b}} | = 2$ fixed, with different values of detuning
frequency: (a) $\delta = 0$ (resonant regime) and (b) $\delta = 6 \kappa_{\ief}$ (nonresonant regime). From these pictures we can
observe that $\textrm{S}(0) = 0$ (this condition refers to separability of the initial density operator for the tripartite system) and
when $\kappa_{\ief} t > 0$, the time-dependent subdeterminant (\ref{e10}) assumes negative values in both situations (the entanglement
condition is promptly veri\-fied). In particular, the inseparability of the reduced density operator turns to be more evident in the
nonresonant regime -- see picture 4(b) -- which coincides with the squeezing effect observed in pictures 2(b) and 2(d) for the variances
$\mathcal{V}_{{\bf Q}}^{({\rm a})}(t)$ and $\mathcal{V}_{{\bf Q}}^{({\rm b})}(t)$, respectively. Indeed, the link between squeezing
effect and entanglement turns more evident through the detuning frequency $\delta = \om_{0} - \om$, this frequency being associated with
the atomic transition $(\om_{0})$ and cavity (driving) field $(\om)$ frequencies in the driven JCM. Thus, by means of the numerical
evidence obtained in this section, we can conclude that $\textrm{S}(t)$ corroborates the Shchukin-Vogel criterion and consequently, it
can be considered the first quantitative characterization of the entanglement properties for the theoretical model under investigation.

\section{Conclusions}

\hp In this paper we have applied the decomposition formula for $\mathfrak{su}(2)$ Lie algebra on the driven Jaynes-Cummings model in
order to calculate the exact analytical expressions for the normally ordered moments when the atom is initially prepared in the excited
state. In particular, adopting the diagonal representation of coherent states, we have shown that the Wigner characteristic function
can be written in the integral form, with their integrands having a commom term which describes the product of the Glauber-Sudarshan
quasiprobability distribution functions for each field, and a kernel responsible for the entanglement. Next, we have evaluated the
normally ordered moments for the cavity and driving fields separately, by means of their respective Wigner characteristic functions. It
is worth emphasizing that the mathematical procedure adopted here does not present any restrictions on the states of the cavity and
external electromagnetic fields. To illustrate our exact results we have fixed, for convenience in the numerical calculations, both the
cavity and driving fields in the coherent states. This procedure has allowed us to investigate not only the squeezing effect associated
with the quadrature operators ${\bf Q}_{{\rm a}({\rm b})}(t)$ and ${\bf P}_{{\rm a}({\rm b})}(t)$ of the cavity (driving) 
electromagnetic field, but also the EPR uncertainty and the Shchukin-Vogel criterion. Such numerical investigation has produced some
important results, within which some deserve to be mentioned: (i) we have shown that squeezing effect is directly related to detuning
frequency in the driven JCM; (ii) the EPR uncertainty has failed in the description of entanglement properties of the system under
consideration; and finally, (iii) we have demonstrated that the Shchukin-Vogel criterion correctly establishes a set of necessary and
sufficient conditions for the negativity of partial transposition of bipartite quantum states. 

\section*{Acknowledgements}

\hp The author is grateful to Di\'{o}genes Galetti and Maurizio Ruzzi from Instituto de F\'{\i}sica Te\'{o}rica (IFT, Unesp, SP, Brazil)
for reading the manuscript and for providing valuable suggestions. This work was supported by Conselho Nacional de Desenvolvimento
Cient\'{\i}fico e Tecnol\'{o}gico (CNPq), Brazil. 

\appendix
\section{Exact analytical expressions for the generalized moments}

\hp The main aim of this Appendix is to obtain an exact analytical expression for the generalized moments 
\be
\lb{a1}
\llg {\bf a}^{\dagger p}(t) {\bf a}^{q}(t) {\bf b}^{\dagger r}(t) {\bf b}^{s}(t) \rg \equiv \Tr \lbk \ro(t) \, {\bf a}^{\dagger p} 
{\bf a}^{q} {\bf b}^{\dagger r} {\bf b}^{s} \rbk 
\ee
with $\{ p,q,r,s \} \in \mathbb{N}$, by means of the Wigner characteristic function
\be
\lb{a2}
\Lambda_{\dab}(\xi_{{\rm a}},\xi_{{\rm a}}^{\ast},\xi_{{\rm b}},\xi_{{\rm b}}^{\ast};t) \equiv \Tr \lbk \ro(t) {\bf D}_{{\rm a}}
(\xi_{{\rm a}}) {\bf D}_{{\rm b}}(\xi_{{\rm b}}) \rbk
\ee
which involves the expectation value of the product of displacement operators for each field in the driven JCM. To this end, let us
establish the link between both expressions through the mathematical relation \cite{r41}
\be
\lb{a3}
\llg {\bf a}^{\dagger p}(t) {\bf a}^{q}(t) {\bf b}^{\dagger r}(t) {\bf b}^{s}(t) \rg = (-1)^{q+s} \frac{\upartial^{p+q+r+s}}
{\upartial \xi_{{\rm a}}^{p} \upartial \xi_{{\rm a}}^{\ast q} \upartial \xi_{{\rm b}}^{r} \upartial \xi_{{\rm b}}^{\ast s}} \exp \lbk \half
\lpar | \xi_{{\rm a}} |^{2} + | \xi_{{\rm b}} |^{2} \rpar \rbk \biggl. \Lambda_{\dab}(\xi_{{\rm a}},\xi_{{\rm a}}^{\ast},\xi_{{\rm b}},
\xi_{{\rm b}}^{\ast};t) \biggr|_{\xi_{{\rm a}},\xi_{{\rm a}}^{\ast},\xi_{{\rm b}},\xi_{{\rm b}}^{\ast} = 0} \; .
\ee
Thus, our first task will consist in obtaining the Wigner characteristic function (\ref{a2}), while the second task will focus upon the
evaluation of the generalized moments (\ref{a1}) with the help of equation (\ref{a3}).

Adopting the same mathematical recipe used in Sec. II to derive Eq. (\ref{e1}), the generalized Wigner characteristic function is
expressed into this context as
\be
\lb{a4}
\Lambda_{\dab}(\xi_{{\rm a}},\xi_{{\rm a}}^{\ast},\xi_{{\rm b}},\xi_{{\rm b}}^{\ast};t) = \int \!\!\!\!\! \int \frac{d^{2} 
\alf_{{\rm a}} d^{2} \alf_{{\rm b}}}{\pi^{2}} \, P_{{\rm a}}(\alf_{{\rm a}}) P_{{\rm b}}(\alf_{{\rm b}})
\widetilde{\mathcal{K}}_{\xi_{{\rm a}({\rm b})},\xi_{{\rm a}({\rm b})}^{\ast}}(\alf_{{\rm a}},\alf_{{\rm b}};t) \; , 
\ee
where $P_{{\rm a}}(\alf_{{\rm a}})$ and $P_{{\rm b}}(\alf_{{\rm b}})$ are the Glauber-Sudarshan quasiprobability distribution functions
associated with the cavity $({\rm a})$ and driving $({\rm b})$ fields, and
\be
\lb{a5}
\widetilde{\mathcal{K}}_{\xi_{{\rm a}({\rm b})},\xi_{{\rm a}({\rm b})}^{\ast}}(\alf_{{\rm a}},\alf_{{\rm b}};t) = \,
_{11}\mathcal{D}_{\xi_{{\rm a}({\rm b})},\xi_{{\rm a}({\rm b})}^{\ast}}^{11}(\alf_{{\rm a}},\alf_{{\rm b}};t) + \,
_{21}\mathcal{D}_{\xi_{{\rm a}({\rm b})},\xi_{{\rm a}({\rm b})}^{\ast}}^{21}(\alf_{{\rm a}},\alf_{{\rm b}};t) 
\ee
represents a kernel responsible for the entanglement between the displacement operators ${\bf D}_{{\rm a}}(\xi_{{\rm a}})$ and 
${\bf D}_{{\rm b}}(\xi_{{\rm b}})$ with
\be
\lb{a6}
_{{\rm ij}}\mathcal{D}_{\xi_{{\rm a}({\rm b})},\xi_{{\rm a}({\rm b})}^{\ast}}^{{\rm ij}}(\alf_{{\rm a}},\alf_{{\rm b}};t) = \llg
\alf_{{\rm a}},\alf_{{\rm b}} | \opu_{{\rm ij}}^{\dagger}(t) {\bf D}_{{\rm a}}(\xi_{{\rm a}}) {\bf D}_{{\rm b}}(\xi_{{\rm b}})
\opu_{{\rm ij}}(t) | \alf_{{\rm a}},\alf_{{\rm b}} \rg \; .
\ee
After lengthy calculations, the analytical expressions for the mean values (\ref{a6}) assume exact forms similar to that obtained for
the cavity field but differ in the dependence on the variables $\xi_{{\rm a}({\rm b})}$ and $\xi_{{\rm a}({\rm b})}^{\ast}$. As a
consequence of this fact, the complex function ${\rm Y}_{\xi,\xi^{\ast}}^{(m,m^{\prime})} (\alf_{{\rm a}},\alf_{{\rm b}})$ must be
adequately substituted by
\brr
\mathcal{H}_{\xi_{{\rm a}({\rm b})},\xi_{{\rm a}({\rm b})}^{\ast}}^{(m,m^{\prime})}(\alf_{{\rm a}},\alf_{{\rm b}}) &=& \exp \lbr -
\half \lpar | \xi_{{\rm a}} |^{2} + | \xi_{{\rm b}} |^{2} \rpar + 2 \im \ima \lbk \lpar \eps_{{\rm b}} \alf_{{\rm a}} - \eps_{{\rm a}}
\alf_{{\rm b}} \rpar^{\ast} \lpar \eps_{{\rm b}} \xi_{{\rm a}} - \eps_{{\rm a}} \xi_{{\rm b}} \rpar \rbk \rbr \nn \\
& & \times \lbk \lpar \eps_{{\rm a}} \alf_{{\rm a}} + \eps_{{\rm b}} \alf_{{\rm b}} \rpar^{\ast} \lpar \eps_{{\rm a}} \xi_{{\rm a}} +
\eps_{{\rm b}} \xi_{{\rm b}} \rpar \rbk^{m^{\prime} - m} L_{m}^{(m^{\prime} - m)} \lpar | \eps_{{\rm a}} \xi_{{\rm a}} + \eps_{{\rm b}}
\xi_{{\rm b}} |^{2} \rpar \nn
\err
for each situation described by the indices i,j= 1,2. In particular, if one consi\-ders the initial states of the cavity and driving
fields in the coherent states, we obtain $\Lambda_{\dab}(\xi_{{\rm a}},\xi_{{\rm a}}^{\ast},\xi_{{\rm b}},\xi_{{\rm b}}^{\ast};t) =
\widetilde{\mathcal{K}}_{\xi_{{\rm a}({\rm b})},\xi_{{\rm a}({\rm b})}^{\ast}}(\nu_{{\rm a}},\nu_{{\rm b}};t)$ and consequently, this
characteristic function assumes the simple form
\be
\lb{a7}
\Lambda_{\dab}(\xi_{{\rm a}},\xi_{{\rm a}}^{\ast},\xi_{{\rm b}},\xi_{{\rm b}}^{\ast};t) = \,
_{11}\mathcal{D}_{\xi_{{\rm a}({\rm b})},\xi_{{\rm a}({\rm b})}^{\ast}}^{11} (\nu_{{\rm a}},\nu_{{\rm b}};t) + \,
_{21}\mathcal{D}_{\xi_{{\rm a}({\rm b})},\xi_{{\rm a}({\rm b})}^{\ast}}^{21} (\nu_{{\rm a}},\nu_{{\rm b}};t) \, .
\ee
Following, we will use (\ref{a4}) to derive an exact analytical expression for the generalized moments (\ref{a3}).

Let us initially introduce the generating function for $\mathcal{H}_{\xi_{{\rm a}({\rm b})},\xi_{{\rm a}({\rm b})}^{\ast}}^{(m, m^{\prime})}(\alf_{{\rm a}},\alf_{{\rm b}})$ through the definition
\be
\lb{a8}
\mathcal{H}_{\xi_{{\rm a}({\rm b})},\xi_{{\rm a}({\rm b})}^{\ast}}^{(m,m^{\prime})}(\alf_{{\rm a}},\alf_{{\rm b}}) = \frac{1}{m!}
\left. \frac{\upartial^{m+m^{\prime}}}{\upartial u^{m} \upartial v^{m^{\prime}}} \, \mathcal{G}_{\xi_{{\rm a}({\rm b})},
\xi_{{\rm a}({\rm b})}^{\ast}}^{(u,v)} (\alf_{{\rm a}},\alf_{{\rm b}}) \right|_{u,v=0}
\ee
where
\be
\lb{a9}
\mathcal{G}_{\xi_{{\rm a}({\rm b})},\xi_{{\rm a}({\rm b})}^{\ast}}^{(u,v)}(\alf_{{\rm a}},\alf_{{\rm b}}) = 
\sum_{k,k^{\prime} = 0}^{\infty} \mathcal{H}_{\xi_{{\rm a}({\rm b})},\xi_{{\rm a}({\rm b})}^{\ast}}^{(k,k^{\prime})}
(\alf_{{\rm a}},\alf_{{\rm b}}) \, \frac{u^{k} v^{k^{\prime}}}{k^{\prime}!} \; .
\ee
Furthermore, we also consider the auxiliar function
\be
\lb{a10}
\mathcal{I}_{m,m^{\prime}}^{(p,q,r,s)}(\alf_{{\rm a}},\alf_{{\rm b}}) = (-1)^{q+s} \frac{\upartial^{p+q+r+s}}{\upartial 
\xi_{{\rm a}}^{p} \upartial \xi_{{\rm a}}^{\ast q} \upartial \xi_{{\rm b}}^{r} \upartial \xi_{{\rm b}}^{\ast s}} \exp \lbk \half \lpar 
| \xi_{{\rm a}} |^{2} + | \xi_{{\rm b}} |^{2} \rpar \rbk \biggl. \mathcal{H}_{\xi_{{\rm a}({\rm b})},\xi_{{\rm a}
({\rm b})}^{\ast}}^{(m,m^{\prime})} (\alf_{{\rm a}},\alf_{{\rm b}}) \biggr|_{\xi_{{\rm a}},\xi_{{\rm a}}^{\ast},\xi_{{\rm b}},
\xi_{{\rm b}}^{\ast} = 0}
\ee
which has a fundamental role in the present context since this function should reduce the complexity of our calculations. Indeed,
substituting (\ref{a8}) into Eq. (\ref{a10}) we obtain promptly the following useful relation: 
\be
\lb{a11}
\mathcal{I}_{m,m^{\prime}}^{(p,q,r,s)}(\alf_{{\rm a}},\alf_{{\rm b}}) = \frac{1}{m!} \left. \frac{\upartial^{m+m^{\prime}}}
{\upartial u^{m} \upartial v^{m^{\prime}}} \, \mathcal{J}_{u,v}^{(p,q,r,s)} (\alf_{{\rm a}},\alf_{{\rm b}}) \right|_{u,v=0}
\ee
with $\mathcal{J}_{u,v}^{(p,q,r,s)}(\alf_{{\rm a}},\alf_{{\rm b}})$ given by
\be
\lb{a12}
\mathcal{J}_{u,v}^{(p,q,r,s)}(\alf_{{\rm a}},\alf_{{\rm b}}) = (-1)^{q+s} \frac{\upartial^{p+q+r+s}}{\upartial \xi_{{\rm a}}^{p}
\upartial \xi_{{\rm a}}^{\ast q} \upartial \xi_{{\rm b}}^{r} \upartial \xi_{{\rm b}}^{\ast s}} \exp \lbk \half \lpar | \xi_{{\rm a}} |^{2} 
+ | \xi_{{\rm b}} |^{2} \rpar \rbk \biggl. \mathcal{G}_{\xi_{{\rm a}({\rm b})},\xi_{{\rm a}({\rm b})}^{\ast}}^{(u,v)}(\alf_{{\rm a}},
\alf_{{\rm b}}) \biggr|_{\xi_{{\rm a}},\xi_{{\rm a}}^{\ast},\xi_{{\rm b}},\xi_{{\rm b}}^{\ast} = 0} \; .
\ee
For instance, the generating function (\ref{a9}) reads as
\brr
\mathcal{G}_{\xi_{{\rm a}({\rm b})},\xi_{{\rm a}({\rm b})}^{\ast}}^{(u,v)}(\alf_{{\rm a}},\alf_{{\rm b}}) &=& \exp \lbr - \half \lpar
| \xi_{{\rm a}} |^{2} + | \xi_{{\rm b}} |^{2} \rpar + 2 \im \ima \lbk \lpar \eps_{{\rm b}} \alf_{{\rm a}} - \eps_{{\rm a}} 
\alf_{{\rm b}} \rpar^{\ast} \lpar \eps_{{\rm b}} \xi_{{\rm a}} - \eps_{{\rm a}} \xi_{{\rm b}} \rpar \rbk \rbr \nn \\
& & \times \exp \lbk - \frac{\lpar \eps_{{\rm a}} \xi_{{\rm a}} + \eps_{{\rm b}} \xi_{{\rm b}} \rpar^{\ast}}{\lpar \eps_{{\rm a}}
\alf_{{\rm a}} + \eps_{{\rm b}} \alf_{{\rm b}} \rpar^{\ast}} \, u + \lpar \eps_{{\rm a}} \alf_{{\rm a}} + \eps_{{\rm b}} \alf_{{\rm b}}
\rpar^{\ast} \lpar \eps_{{\rm a}} \xi_{{\rm a}} + \eps_{{\rm b}} \xi_{{\rm b}} \rpar v + uv \rbk \nn
\err
while (\ref{a12}) has the explicit expression
\brr
\mathcal{J}_{u,v}^{(p,q,r,s)}(\alf_{{\rm a}},\alf_{{\rm b}}) &=& \lbk \eps_{{\rm a}} \lpar \eps_{{\rm a}} \alf_{{\rm a}} + 
\eps_{{\rm b}} \alf_{{\rm b}} \rpar^{\ast} v + \eps_{{\rm b}} \lpar \eps_{{\rm b}} \alf_{{\rm a}} - \eps_{{\rm a}} \alf_{{\rm b}}
\rpar^{\ast} \rbk^{p} \lbk \frac{\eps_{{\rm a}} u}{\lpar \eps_{{\rm a}} \alf_{{\rm a}} + \eps_{{\rm b}} \alf_{{\rm b}} \rpar^{\ast}} + 
\eps_{{\rm b}} \lpar \eps_{{\rm b}} \alf_{{\rm a}} - \eps_{{\rm a}} \alf_{{\rm b}} \rpar \rbk^{q} \nn \\
& & \times \lbk \eps_{{\rm b}} \lpar \eps_{{\rm a}} \alf_{{\rm a}} + \eps_{{\rm b}} \alf_{{\rm b}} \rpar^{\ast} v - \eps_{{\rm a}}
\lpar \eps_{{\rm b}} \alf_{{\rm a}} - \eps_{{\rm a}} \alf_{{\rm b}} \rpar^{\ast} \rbk^{r} \lbk \frac{\eps_{{\rm b}} u}{\lpar 
\eps_{{\rm a}} \alf_{{\rm a}} + \eps_{{\rm b}} \alf_{{\rm b}} \rpar^{\ast}} - \eps_{{\rm a}} \lpar \eps_{{\rm b}} \alf_{{\rm a}} -
\eps_{{\rm a}} \alf_{{\rm b}} \rpar \rbk^{s} \exp (uv) \; . \nn 
\err
Consequently, the partial derivatives of this expression with respect to $u$ and $v$ at the point $u,v=0$ allow us to determine the
auxiliar function (\ref{a11}). 

Finally, it is worth mentioning that the generalized moments for any kind of cavity and driving fields can be obtained by means of a
basic sequence of functions evaluated in this appendix -- e.g., the schematic diagram showed below
\bd
\xymatrix{
& \mathcal{H}_{\xi_{{\rm a}({\rm b})},\xi_{{\rm a}({\rm b})}^{\ast}}^{(m,m^{\prime})}(\alf_{{\rm a}},\alf_{{\rm b}}) \ar@{|->}[dddd]
\ar@{<->}[rd] & \\
_{ij}\mathcal{D}_{\xi_{{\rm a}({\rm b})},\xi_{{\rm a}({\rm b})}^{\ast}}^{kl}(\alf_{{\rm a}},\alf_{{\rm b}};t) \ar@{|->}[ru] & &
\mathcal{G}_{\xi_{{\rm a}({\rm b})}, \xi_{{\rm a}({\rm b})}^{\ast}}^{(u ,v)}(\alf_{{\rm a}},\alf_{{\rm b}}) \ar@{|-->}[d] \\
\widetilde{\mathcal{K}}_{\xi_{{\rm a}({\rm b})},\xi_{{\rm a}({\rm b})}^{\ast}}(\alf_{{\rm a}},\alf_{{\rm b}};t) \ar@{|->}[u] & & 
\mathcal{J}_{u,v}^{(p,q,r,s)}(\alf_{{\rm a}},\alf_{{\rm b}}) \ar@{|-->}[d] \\
\Lambda_{\dab}(\xi_{{\rm a}},\xi_{{\rm a}}^{\ast},\xi_{{\rm b}},\xi_{{\rm b}}^{\ast};t) \ar@{|->}[u] & &
\mathcal{I}_{m,m^{\prime}}^{(p,q,r,s)}(\alf_{{\rm a}},\alf_{{\rm b}}) \ar@{|-->}[ld] \\
       & \llg {\bf a}^{\dagger p}(t) {\bf a}^{q}(t) {\bf b}^{\dagger r}(t) {\bf b}^{s}(t) \rg \ar@{|->}[lu] & }
\ed
represents two alternative ways of finding out the generalized moments (\ref{a1}). From the theoretical point of view, (\ref{a1}) plays
an essential role not only in the investigation process on both the squeezing effect and the EPR uncertainty, but also in the analysis
of some inseparability criteria for the driven JCM.


\end{document}